%% file: revised01.tex
\documentclass[preprint,12pt]{elsarticle}
\usepackage{hyperref}
\usepackage{xurl}
\usepackage{algorithm}%
\usepackage{algorithmicx}%
\usepackage{algpseudocode}
\usepackage{manyfoot}%
\usepackage{booktabs}%
\usepackage{caption}%
\usepackage{multirow}
\usepackage{glossaries}
\usepackage{rotating}
\usepackage{float}
\usepackage{threeparttable}
\usepackage{soul}
\usepackage[dvipsnames]{xcolor}
\usepackage{amsmath,amssymb,amsfonts}%
\usepackage{amsthm}%
\usepackage{mathrsfs}%
\input{glossary}

\journal{Journal of Voice}

\begin{document}

\begin{frontmatter}



\title{Reproducible Machine Learning-based Voice Pathology Detection: Introducing the Pitch Difference Feature}


\author[1,2]{Jan Vrba\fnref{eq}} 
\ead{jan.vrba@vscht.cz}
\author[1,2]{Jakub Steinbach\fnref{eq}} 
\ead{jakub.steinbach@vscht.cz}
\author[1]{Tomáš Jirsa\fnref{eq}} 
\ead{tomas.jirsa@vscht.cz}
\author[3]{Laura Verde} 
\ead{laura.verde@unicampania.it}
\author[3]{Roberta De Fazio} 
\ead{roberta.defazio@unicampania.it}
\author[2]{Yuwen Zeng} 
\ead{yuwen@tohoku.ac.jp}
\author[2]{Kei Ichiji} 
\ead{ichiji@tohoku.ac.jp}
\author[1]{Lukáš Hájek} 
\ead{lukas.hajek@vscht.cz}
\author[1]{Zuzana Sedláková} 
\ead{zuzana2.sedlakova@vscht.cz}
\author[4]{Zuzana Urbániová} 
\ead{zuzana.urbaniova@fnkv.cz}
\author[4]{Martin Chovanec} 
\ead{martin.chovanec@fnkv.cz}
\author[1]{Jan Mareš} 
\ead{jan.mares@vscht.cz}
\author[2]{Noriyasu Homma} 
\ead{homma@tohoku.ac.jp}

\fntext[eq]{These authors contributed equally to this work.}

\affiliation[1]{organization={Department of Mathematics, Informatics, and Cybernetics, University of Chemistry and Technology, Prague}, 
            addressline={Technická 5}, 
            city={Prague},
            postcode={166 28}, 
            country={Czech Republic}}

\affiliation[2]{organization={Department of Radiological Imaging and Informatics, Tohoku University Graduate School of Medicine}, 
            addressline={2-1-1 Katahira, Aoba-ku}, 
            city={Sendai},
            postcode={980-8577}, 
            country={Japan}}

\affiliation[3]{organization={Department of Mathematics and Physics, University of Campania ”Luigi Vanvitelli”}, 
            addressline={Viale Abramo Lincoln 5}, 
            city={Caserta},
            postcode={81100}, 
            country={Italy}}

\affiliation[4]{organization={Department of Otorhinolaryngology, Faculty Hospital Královské Vinohrady}, 
            addressline={Šrobárova 1150/50}, 
            city={Prague},
            postcode={100 34}, 
            country={Czech Republic}}
\begin{abstract}
\textbf{Purpose:} 
We introduce a novel methodology for voice pathology detection using the publicly available Saarbrücken Voice Database (SVD) and a robust feature set combining commonly used acoustic handcrafted features with two novel ones: pitch difference (relative variation in fundamental frequency) and NaN feature (failed fundamental frequency estimation).

\textbf{Methods:}
We evaluate six machine learning (ML) algorithms --- support vector machine, k-nearest neighbors, naive Bayes, decision tree, random forest, and AdaBoost --- using grid search for feasible hyperparameters and 20480 different feature subsets. Top 1000 classification models -- feature subset combinations for each ML algorithm are validated with repeated stratified cross-validation. To address class imbalance, we apply K-Means SMOTE to augment the training data.

\textbf{Results:}
Our approach achieves 85.61\%, 84.69\% and 85.22\% unweighted average recall (UAR) for females, males and combined results respectively. We intentionally omit accuracy as it is a highly biased metric for imbalanced data. 

\textbf{Conclusion:}
Our study demonstrates that by following the proposed methodology and feature engineering, there is a potential in detection of various voice pathologies using ML models applied to the simplest vocal task, a sustained utterance of the vowel /a:/.
To enable easier use of our methodology and to support our claims, we provide a publicly available GitHub repository with DOI \href{https://doi.org/10.5281/zenodo.13771573}{10.5281/zenodo.13771573}. Finally, we provide a REFORMS checklist to enhance readability, reproducibility and justification of our approach.
\end{abstract}

\begin{graphicalabstract}
\includegraphics[width=\textwidth]{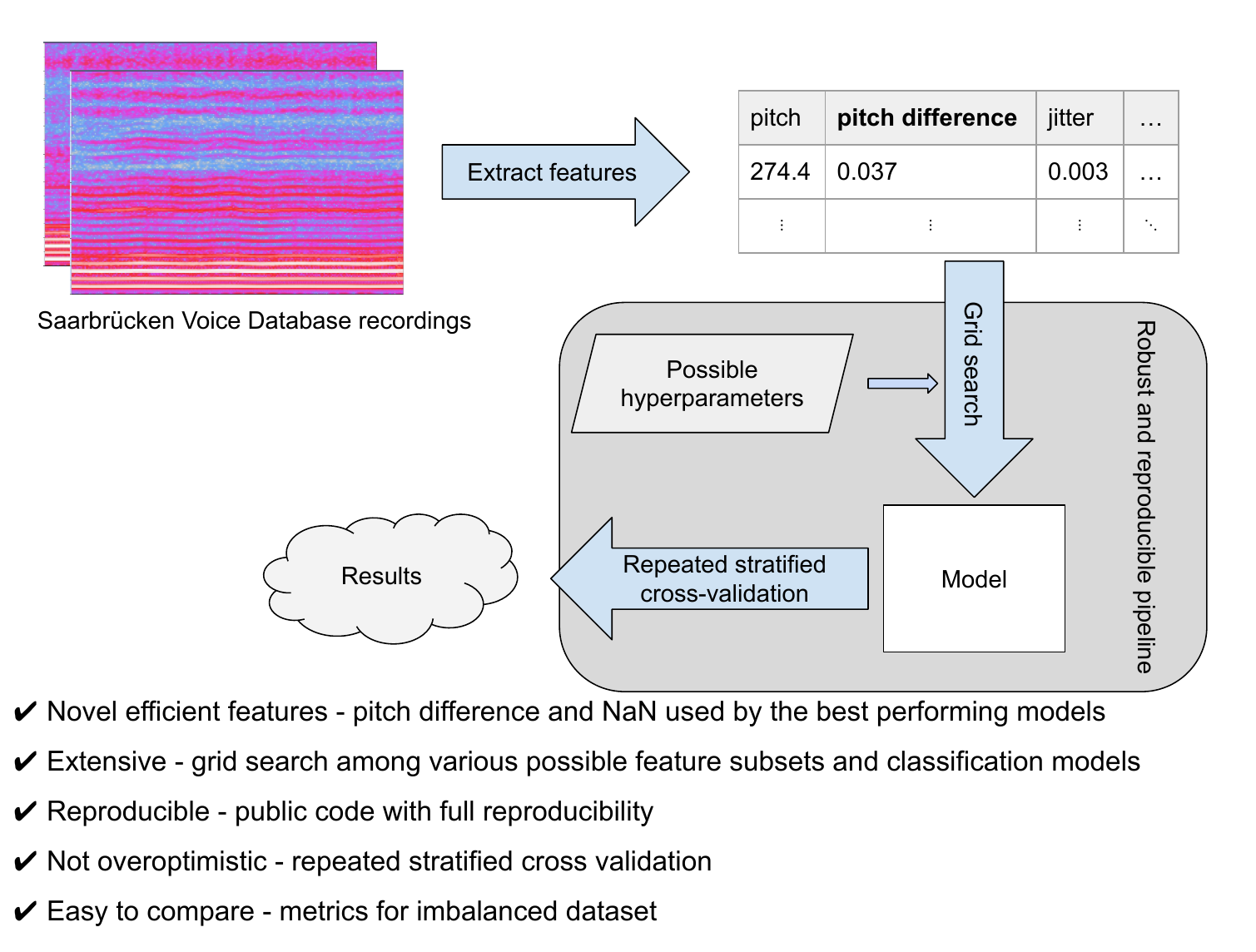}
\end{graphicalabstract}

\begin{highlights}
\item Introducing reproducible machine-learning-based system for voice pathology detection.
\item Novel pitch difference and NaN features improving classification performance.
\item Addressing class imbalance with k-means SMOTE, boosting minority class predictions.
\item Avoiding data leakage by managing multiple recordings per patient effectively.
\item Performing extensive grid search for optimal features and hyperparameters.
\end{highlights}


\end{frontmatter}



\section{Introduction}\label{sec:introduction}
Voice and speech are fundamental aspects of human communication, playing a crucial role in social interaction, emotional expression, and professional performance. Disorders affecting these functions can have a profound impact on an individual’s quality of life, leading to reduced intelligibility, impaired social interactions, and psychological distress. Voice and speech pathologies encompass a diverse range of conditions, each characterized by distinct alterations in the acoustic and articulatory properties of speech signals. A precise understanding of these pathologies is essential for accurate diagnosis, effective therapeutic intervention, and improved clinical outcomes. \cite{stachler2018clinical}

Traditionally, voice disorders have often been described primarily in terms of voice quality disturbances, as seen in conditions such as dysphonia, which is commonly associated with hoarseness, breathiness, or strain. However, this perspective is too narrow to encompass the full spectrum of speech and voice disorders. For instance, dysarthria, a motor speech disorder resulting from neurological impairment, affects articulation, prosody, resonance, and speech duration, rather than merely altering voice quality. Similarly, conditions such as apraxia of speech and neurological voice disorders, including spasmodic dysphonia and hypokinetic dysarthria due to the Parkinson's disease, lead to a combination of deficits in phonation, speech timing, and intonation. \cite{illner2023automated, portalete2024acoustic}
Moreover, emerging research has emphasized the importance of objective acoustic and articulatory analyses in the evaluation of voice and speech disorders. 

Advancements in computerized speech analysis, laryngeal imaging, and neurophysiological assessments have provided deeper insights into the pathophysiological mechanisms underlying these conditions. These tools enable clinicians and researchers to characterize speech disorders more precisely, facilitating early diagnosis and the development of targeted interventions.

This involves the processing and analysis of various acoustic features of the voice signal, which can reveal changes or alterations in voice quality caused by specific diseases \cite{hegde2019survey}.
In this context, artificial intelligence can be a valid and powerful tool as it may support the decision making process in clinical settings. However, utilization of these solutions imposes many additional tasks, ranging from data collection and preprocessing to ground truth labeling and correct algorithm selection.

Moreover, the choice of appropriate acoustic features is fundamental. The use of acoustic features to characterize pathological voice quality has been investigated in a variety of contexts and for a variety of purposes \cite{borsky2017modal,lopes2022performance}. These features can provide a quantitative method of assessing voice characteristics that are otherwise difficult to measure. However, there is no standardized set of acoustic measures, making the selection of appropriate acoustic metrics and their interpretation an ongoing challenge.

In this paper on voice pathology detection, we introduce a methodology for \gls{svd} that effectively addresses the challenge of multiple recordings from the same patients. Without proper handling, such data redundancy can lead to data leakage, an undesirable effect in which the classification model obtains information about the validation set during the training phase, which might lead to overly optimistic classification results \cite{KAPOOR2023100804}. We consider this an important contribution, as this issue, in the context of voice pathology detection, is rarely discussed in the literature.

Additionally, we present two novel features --- pitch difference and NaN feature. Several used features are based on the \gls{f0}. The NaN feature reflects the fact, that we were not able to extract \gls{f0} values in the analyzed speech signal. The pitch difference feature quantifies the variability of \gls{f0} by measuring fluctuations within a single sustained vowel production. We assume that this variability is often altered in pathological voices due to impairments in vocal fold function, making it a possibly useful feature for voice pathology detection.  We experimentally verify proposed features' usefulness in voice pathology detection.

Based on our experimental findings, we emphasize the necessity of training models separately for male and female patients. Given that we do not investigate the causality or direct correlation of individual features with specific pathologies --- and instead work with a large feature set --- we leverage \gls{ml} algorithms. Our feature selection process reveals that the optimal feature sets may vary depending on both the sex and \gls{ml} algorithm. To identify robust feature sets, we train multiple classification models across different combinations of feature subsets (see Table~\ref{tab:features_desc}).

By computing the mean \gls{mcc}, we identify the most promising feature -- classification model combinations. We then perform repeated stratified 10-fold cross-validation for the top-performing models to estimate the average performance metrics along with their standard deviations, providing insight into the confidence of our results. 

To our best knowledge, our work is the first in the field of voice pathology detection to provide fully reproducible results. The code used for computations and feature extraction is available in our repository, ensuring complete transparency and allowing anyone to verify, build upon, or extend our work. By prioritizing reproducibility, we aim to set a standard for rigorous and open research.

\section{Related Works}
\label{sec:related}
In this chapter, we outline the original contribution related to voice pathology detection present in the literature, limiting our research to works connected to \gls{svd}. This choice is due to its comprehensiveness:  it is the only voice recording database representing the common voice pathologies. See Section~\ref{sec:data} for detailed information.

Harar et al. \cite{harar2020towards} test various acoustic features, such as pitch, jitter, shimmer, \gls{hnr}, detrended fluctuation analysis parameters, glottis quotients (open, closed), glottal-to-noise excitation ratio, Teager–Kaiser energy operator, modulation energy, and normalized noise energy as well as \gls{mfcc}. Moreover, they considered the sound samples and their spectrograms as input, all in combination with XGBoost, IsolationForest, and DenseNet models to determine the pathologic samples, reaching an F1 score of 73.3\% and \gls{uar} we computed as 73.3\%. While, in Gupta et al.\cite{10.1109/ICASSP48485.2024.10446075}, features derived from self-supervised learning models, Data2Vec and Wav2Vec, alongside \gls{mfcc} are explored. The reliability of these features to evaluate voice quality is tested using \gls{svm} and \gls{dnn}, achieving an accuracy of 77.83\% and \gls{uar} we computed as 77.86\%.

A different approach is taken by Verde et al.\cite{10.1145/3433993}, where the authors transform sound wave data into spectrograms and treat classification as an image recognition problem using a \gls{cnn}, achieving 73.93\% accuracy and \gls{uar} we computed as 70.68\%.

In Kotarba \& Kotarba\cite{10.21008/j.0860-6897.2021.1.08}, the authors use \gls{mfcc} and \gls{gfcc} with a classification model based on a \gls{nn}, achieving 81.84\% accuracy. Unlike other studies that use the /a:/ sound for feature extraction, they employ whole sentences. Unfortunately, we were not able to compute \gls{uar}, because the reported results appear inconsistent, making it impossible to calculate the metric.
Another study by Harar et al.\cite{10.1109/IWOBI.2017.7985525} tests a \gls{dnn} with convolutional layers, using the voice signal as a feature set for the convolutional layer, achieving 68.08\% accuracy and \gls{uar} we computed as 72.32\%.
Park et al.\cite{10.1109/LSP.2023.3298532} combine \gls{cnn}-based feature extractors with various gls{ml} algorithms, such as \gls{svm} and \gls{dnn}s, achieving an \gls{uar} score of 84.97\% on the entire \gls{svd} dataset. 

Finally, Omeroglu et al.\cite{10.1016/j.jestch.2022.101148} test a combination of \gls{svm} and \gls{cnn}-based feature extractors on spectrograms of sound and \gls{egg} signals. They integrate these with traditional acoustic features like \gls{mfcc}, \gls{lpc}, and \gls{f0}, reaching an accuracy of 90.10\% and \gls{uar} we computed as 88.75\%. Similarly, they reach an accuracy of 87.41\% by extracting the mentioned features and using \gls{svm} for classification. However, they reach such high performance without specifying how they handled multiple recordings from the same patient which occur in the dataset, potentially introducing data leakage and therefore reporting an overly optimistic performance.

In another study by Verde et al.\cite{10.1109/ACCESS.2019.2938265}, the authors take age and sex information from the database and extract various acoustic features from the time domain, such as \gls{f0}, jitter, shimmer, and \gls{hnr}. They employ \gls{ml} algorithms, including boosted trees, \gls{svm}, \gls{dt}, \gls{nb}, and \gls{knn}, achieving the highest accuracy of 84.5\% with a boosted tree model on an imbalanced dataset and \gls{uar} we computed as 84.55\%. In the follow-up study \cite{10.1109/ACCESS.2018.2816338}, they expand their work by incorporating \gls{mfcc} and their derivatives, and test additional \gls{ml} algorithms such as logistic model tree and instance-based learning algorithms, reaching 85.77\% accuracy with a \gls{svm} model and \gls{uar} we computed as 85.77\%. However, they utilized a balanced subset of \gls{svd}.

Additionally, among the researched works, there were several works that took a subset of the \gls{svd} based on selected pathologies.

Tirronen et al.\cite{tirronen2023hierarchical} extract features based on \gls{mfcc} and self-supervised algorithms from samples of healthy individuals and patients with hyperfunctional dysphonia and vocal fold paresis. Using \gls{svm}, they achieve 75.65\% and 74.50\% accuracy for male and female patients, respectively. 
Compared to that, Yagnavajjula et al.\cite{yagnavajjula2024automatic} focus on developing classification models to distinguish between healthy subjects and patients suffering from spasmodic dysphonia and laryngeal nerve paralysis. Their exploration of multi-modal classification methods results in an accuracy of 68.11\%.

Further expanding the scope, Junior et al.\cite{junior2023multiple} investigate multi-modal classification for patients with various conditions, including dysphonia, laryngitis, Reinke’s edema, vox senilis, and central laryngeal motion disorder, using energy, zero-crossing rate, and entropy as features. By employing \gls{svm} and \gls{nn}, they achieve an average accuracy of 88.46\%. 

Similarly, Fan et al.\cite{fan2021class} explore multimodal classification for conditions such as nodules, polyps, edema, and paralysis, as well as binary classification between healthy and unhealthy individuals. They utilize \gls{mfcc} as features and a \gls{smote}-based method for balancing datasets and reach a maximum F1-score of 90\% with a \gls{cnn}-based model for binary classification and we computed \gls{uar} as 90\%. However, they do not describe how they handle duplicities in \gls{svd} dataset and very likely introduce the data leakage due their methodology.

Ding et al.\cite{ding2021deep} propose their own approach by combining \gls{mfcc} and log-mel-frequency spectral coefficients with deep models, using recordings from the \gls{svd} and their own database. This approach results in an accuracy of up to 81.6\%.
Meanwhile, Guedes et al.\cite{guedes2019transfer} focus on patients with dysphonia, chronic laryngitis, and vocal cord paralysis, employing features extracted by a VGGish model in combination with a \gls{lstm} network. They achieve an F1-score of up to 80\% in distinguishing between healthy individuals and those with paralysis.

Hemmerling\cite{hemmerling2017voice} tests quantitative voice parameters combined with a \gls{mlp} model to classify healthy individuals and those with hyperfunctional dysphonia, laryngitis, and recurrent laryngeal nerve paralysis, achieving 87.5\% accuracy. 

Additionally, AL-Dhief et al.\cite{al2023voice} select 280 samples for pathology detection, extracting \gls{mfcc} features and using them in a \gls{dt} model, which results in an accuracy of 67.9\%.

Expanding the feature set, AnilKumar \& Reddy\cite{anilkumar2023classification} use \gls{mfcc}, first and second derivatives of \gls{mfcc}, linear prediction cepstral coefficients, and constant - Q cepstral coefficients with a Bi-\gls{lstm} to classify eight selected pathologies, namely dysody, dysphonia, functional dysphonia, hyperfunctional dysphonia, hypofunctional dysphonia, spasmodic dysphonia, vocal cyst polyp, and healthy individuals, achieving an accuracy of 92.7\%, which is the only metric they provide.
Finally, Tirronen et al.\cite{tirronen2022effect} examine the impact of \gls{mfcc} extraction window length on detecting pathologies in patients with dysphonia and reflux laryngitis. Using \gls{svm}, they achieve up to 75.1\% accuracy and \gls{uar} we computed as 75\%.

Finally, multiple works study the use of end-to-end models, which are not solely dependent on hand-crafted features. \cite{rs-msconvnet,narendra2020glottal} 
Liu et al. \cite{liu2023end} propose an end-to-end deep learning model for classification of laryngitis and hyperfunctional dysphonia using stacked vowels, while reaching \gls{uar} of 72\%.
Reddy et al. \cite{Reddy20221863} apply a wavelet scattering network to extract features from recordings in combination with \gls{mlp} for classification. With their end-to-end approach, they reach 81.32\% \gls{uar} using recordings of sentences as their input.

\section{Materials \& Methods}
\begin{figure*}[h]
    \centering
    \includegraphics[width=1\linewidth]{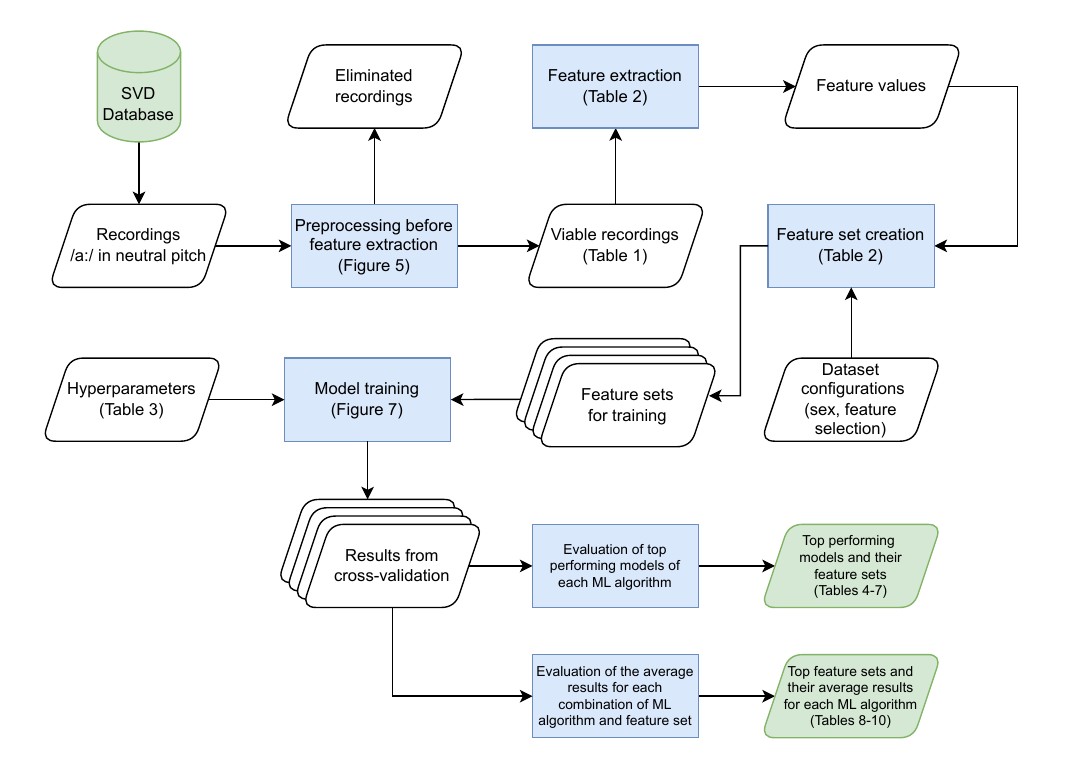}
    \caption{Flow diagram of the proposed methodology}
    \label{fig:flow_overview}
\end{figure*}

To make our methodology more comprehensive, we provide a flowchart to show the process of feature extraction, training and results validation in \autoref{fig:flow_overview}, along with the reference to the relevant tables and figures in the manuscript.

\subsection{Data}\label{sec:data}
Voice pathology detection studies often use the MEEI \cite{meei}, VOICED \cite{voiced, voiceddb}, FEMH Voice Data Challenge 2018 \cite{femh_challenge}, AVPD \cite{avpd} datasets, and several works use their own datasets. However, the MEEI dataset is no longer available, and the FEMH Voice Data Challenge 2018 is not publicly available. We do not consider the AVPD dataset feasible for our study, as it includes only five types of pathologies: vocal fold cysts, nodules, paralysis, polyps, and sulcus. Thus, it does not reflect the number of various pathologies presented in the general population. The VOICED database contains 3 different pathologies, comprising 21 various disorders, but contains a limited sample size of 208 recordings.

Therefore, we use the \gls{svd} \cite{svd} for our work as it is publicly available, contains a wide range of pathologies, and is the largest of the databases. The \gls{svd} was developed by the Phonetics group at the Department of Language Science and Technology, Saarland University, and is available at \url{https://stimmdb.coli.uni-saarland.de}. 

\begin{figure}[h]
    \centering
    \includegraphics[width=0.4\linewidth]{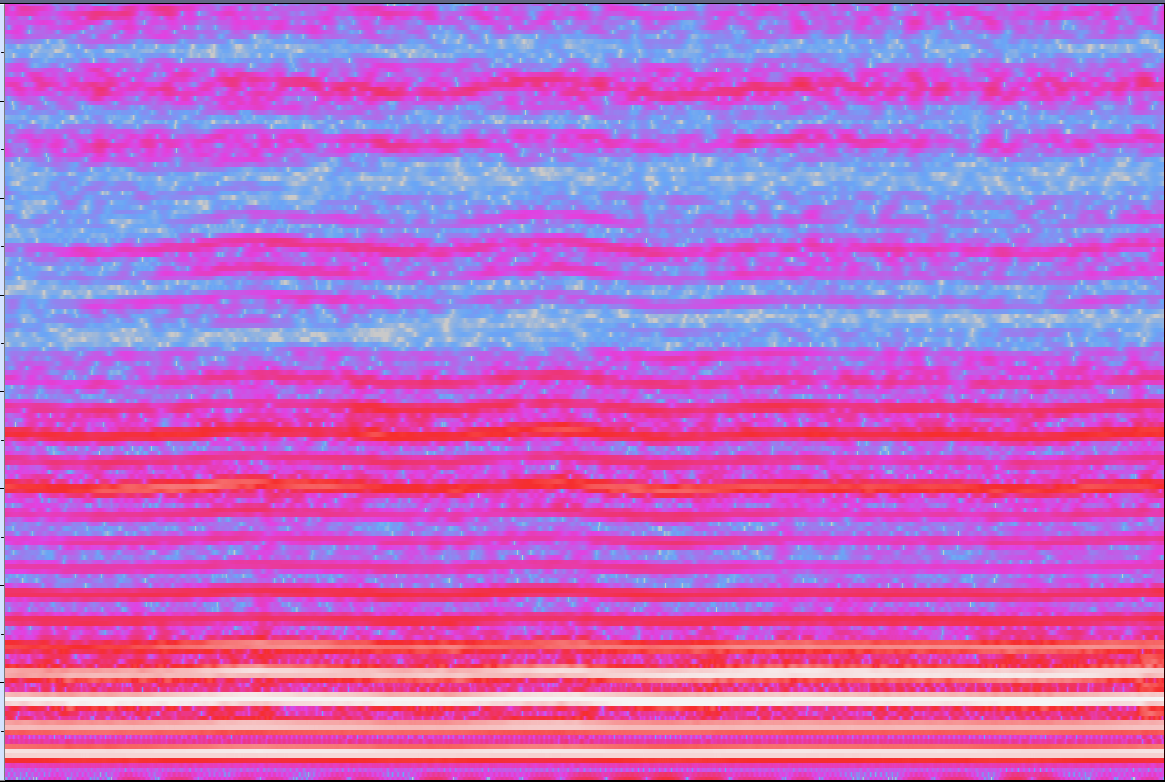}
    \caption{Healthy female subject ID 1}
    \label{subfig:spectrogram-healthy}
\end{figure}%
\begin{figure}[h]
    \centering
    \includegraphics[width=0.4\linewidth]{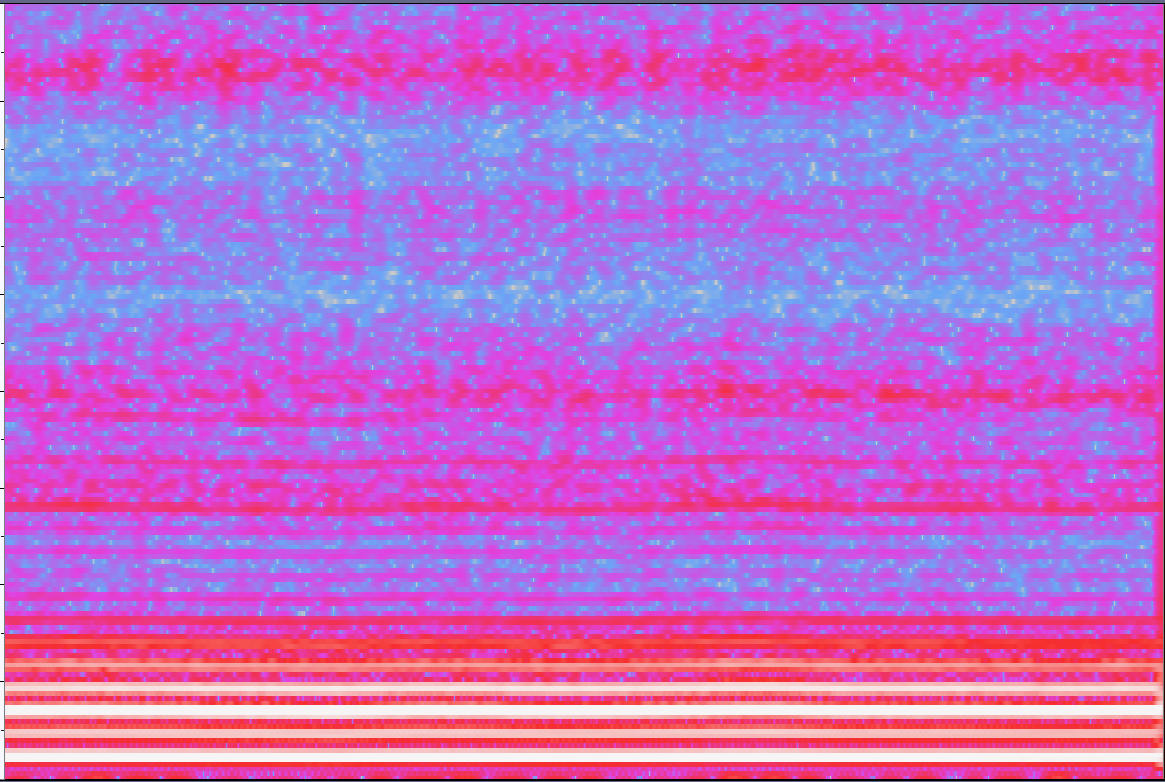}
    \caption{Female subject ID 1, suffering with functional dysphonia}
    \label{subfig:spectrogram-pathological}
\end{figure}

The dataset contains data from 1853 patients and includes voice and \gls{egg} recordings of various vowels in different pitches, as well as from a German phrase, possibly from multiple recording sessions. For each recording session, information concerning the sex and age of the speaker is also provided, as well as a list of pathologies and diagnoses and any comments.

For our research, we use the recordings containing the sound /a:/ in neutral pitch for pathology detection due to its frequent use in previous studies and clinical protocols \cite{dibazar2006pathological, sifel}. Figure \ref{subfig:spectrogram-healthy} and Figure \ref{subfig:spectrogram-pathological} present examples of healthy and pathological spectrograms, illustrating the /a:/ sound in neutral pitch from a healthy female subject (patient ID 1) and the same subject with functional dysphonia. We fully recognize that many voice as well as  speech pathologies go beyond phonatory function and involve articulatory, prosodic, and temporal characteristics, which are more effectively captured in continuous speech.

For feature extraction, we consider some limitations to ensure robust and unbiased results in voice pathology detection. The database contains samples from underage patients. Due to developmental changes at a young age, we exclude any recordings of patients under 18 years of age. Research \cite{BERGER2019580.e21, HOLLIEN2012e29} shows that during development, the characteristics of young voices, such as the fundamental frequency, are distinct from those of fully developed voices. These developmental changes could cause problems for the classification model and reduce its ability to accurately detect pathology. By excluding these age groups, we aim to maintain a more consistent and reliable dataset for our analysis.

Another significant issue is the presence of multiple recordings for some patients. For example, the patient with the talker ID 2027 has 24 recordings of the sound /a:/ in neutral pitch. This repetition poses a risk of data leakage, which can affect the results if not addressed properly.

To prevent this risk in a deterministic way, we select the oldest sample of each type by date and recording ID, resulting in a maximum of two recordings per patient, one healthy and one pathological. We believe that this approach minimizes the likelihood that the classification model learns patient identities, as the patient's state remains independent of their identity.

In addition, we exclude recordings labeled \textit{1573-a-n.wav} and \textit{87-a-n.wav}. The former contains two distinct sound recordings, while the latter is corrupted by an artifact, likely caused by hardware or software errors. Moreover, we exclude several recordings based on the information provided in the comments from the database, stating they contain some artifacts or other problems. Finally, for several recordings, the "Pathologies" column contains information that the speaker performs the tasks during the recording using their singing voice or that they are a singer, which may lead to incorrect labeling of the subject as pathological. We exclude these samples from the study as well.

Next, we trim remaining recordings to eliminate any potential silent parts. Using the \textit{librosa} library \cite{librosa}, we trim the leading and trailing parts of the recordings that are 15 dB quieter than the maximum root mean square value of the amplitude in the analyzed recording.

\begin{figure}[!h]
    \centering
    \includegraphics[width=0.7\linewidth]{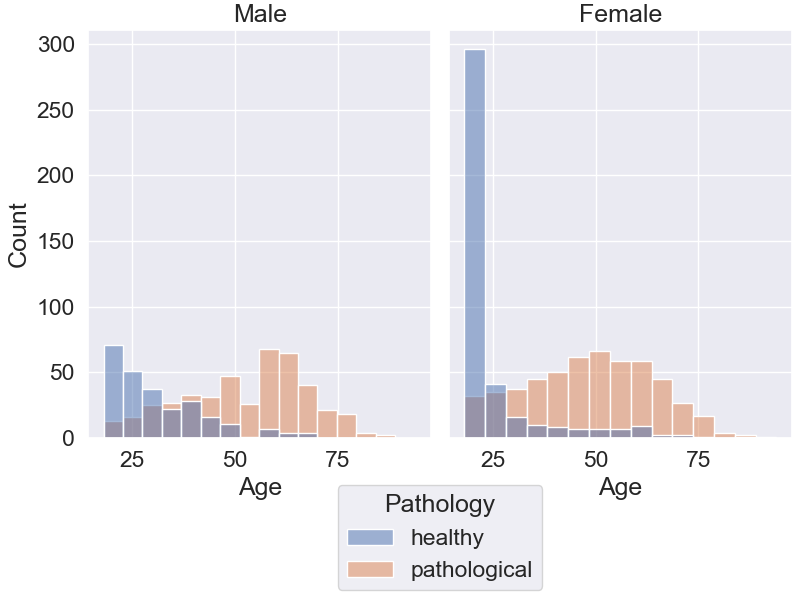}
    \caption{Age distribution of healthy and pathological male and female subjects}
    \label{fig:age}
\end{figure}

The preprocessing step results in 1636 recordings. The age distribution of the data divided by sex and pathology is described in Table~\ref{tab:age_stats} as well as in Figure~\ref{fig:age}. The complete list of recordings excluded from the experiment, along with the reason for their exclusion, is included in the Supplementary Material. The outcome variable takes the value of 0 if no pathology is diagnosed, and the value of 1 if one or more pathologies are diagnosed.

\begin{table*}
\centering
\caption{Age distribution among the sex and voice condition in the used data}
\begin{tabular}{lccccc}
\toprule
 & \multicolumn{2}{c}{Female} & \multicolumn{2}{c}{Male} & \multirow{2}{*}{Unit}\\
& Healthy & Pathological & Healthy & Pathological &\\
\midrule
Mean & 25.38 & 48.55 & 31.45 & 52.38 & \multirow{7}{*}{\rotatebox[origin=c]{90}{years}}  \\
Standard deviation & 11.21 & 15.28 & 11.52 & 15.12 & \\
Minimum & 18.00 & 18.00 & 18.00 & 18.00 & \\
25\% percentile & 20.00 & 36.00 & 22.00 & 41.00 & \\
50\% percentile & 21.00 & 49.00 & 28.00 & 55.50 & \\
75\% percentile & 24.00 & 60.00 & 38.00 & 63.00 & \\
Maximum & 84.00 & 94.00 & 69.00 & 89.00 & \\
\midrule
Total number of subjects & 407 & 541 & 252 & 436  & - \\
\bottomrule
\end{tabular}
\label{tab:age_stats}
\end{table*}

\begin{figure}[!h]
    \centering
    \includegraphics[width=0.6\linewidth]{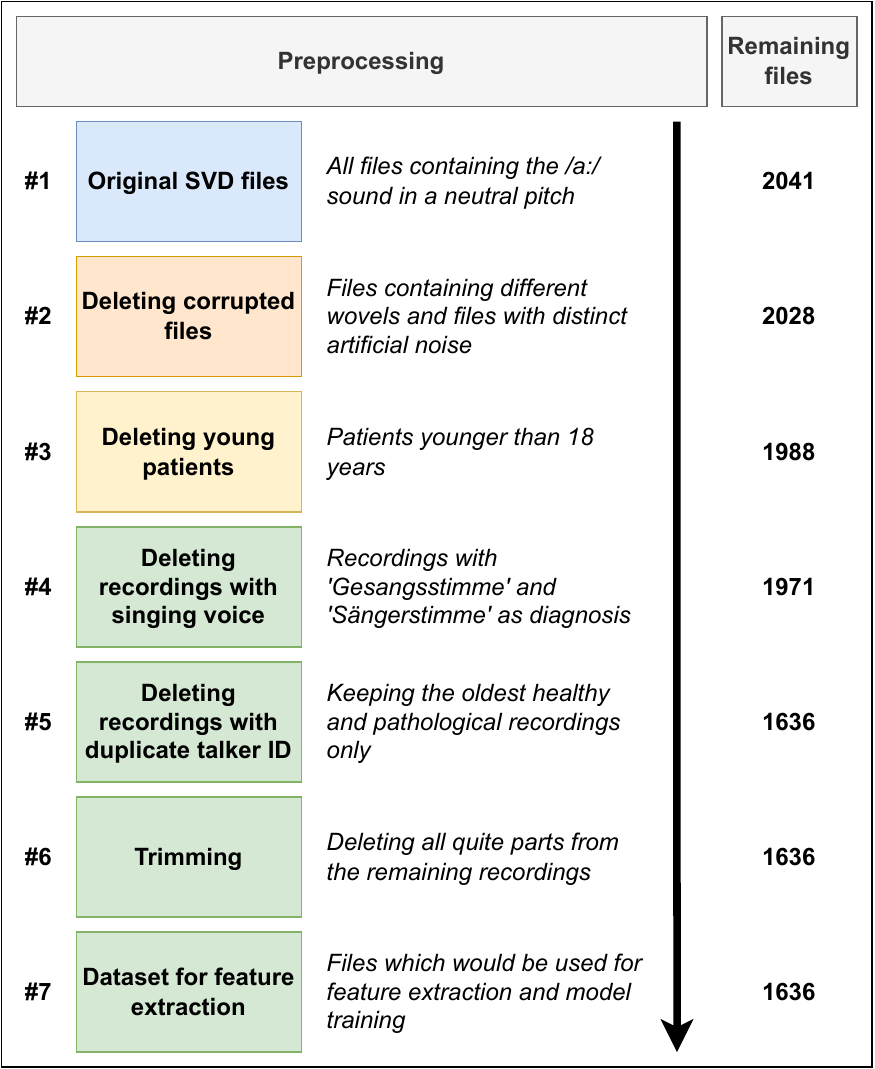}
    \caption{Preprocessing before feature extraction}
    \label{fig:flow_prep}
\end{figure}

The whole preprocessing workflow is illustrated in Figure~\ref{fig:flow_prep}.

\subsection{Feature Extraction} 
\label{sec:feature}
Acoustic features we use can be categorized into time domain, spectral, and cepstral features. We rely on various Python libraries for the feature extraction, namely \textit{parselmouth} \cite{parselmouth} for features related to \gls{f0} and formants, \textit{spkit} \cite{nikesh_bajaj_2021_4710694} for Shannon entropy, \textit{torchaudio} \cite{torchaudio} for \gls{lfcc}, \textit{librosa} for the remaining acoustic features, and \textit{SciPy} \cite{2020SciPy-NMeth} and \textit{NumpPy} \cite{2020NumPy-Array} to calculate statistical values from the extracted features or from the signal. Table~\ref{tab:features_desc} describes libraries used to extract each feature. It also contains the list of all features used in this article, along with references to other related papers utilizing these features.

In addition to the aforementioned features, we also included information about the age of the speakers, as age has an effect on the overall quality of the voice and because this data is practically always known.

\begin{sidewaystable*}
\centering
\small
\caption{Extracted features, their notation, their use in feature sets, and library used for extraction}
\label{tab:features_desc}
\begin{tabular}{{p{4cm}cccc}}
\toprule
Feature                                                 &Symbol             &Configuration for feature set creation &Python library                     & Used in articles \\
\midrule
mean \gls{f0} across all window of the signal           &$\overline{f}_{0}$ &used in all feature sets               &parselmouth \cite{parselmouth}     &\cite{ lopes2022performance,harar2020towards,10.1016/j.jestch.2022.101148,10.1109/ACCESS.2019.2938265,10.1109/ACCESS.2018.2816338}\\
harmonic-to-noise ratio                                 &$\overline{HNR}$   &used in all feature sets               &parselmouth \cite{parselmouth}     &\cite{ harar2020towards,10.1109/ACCESS.2019.2938265,10.1109/ACCESS.2018.2816338}\\
jitter                                                  &$jitta$            &used in all feature sets               &parselmouth \cite{parselmouth}     &\cite{ lopes2022performance,harar2020towards,10.1109/ACCESS.2019.2938265,10.1109/ACCESS.2018.2816338}\\
shimmer                                                 &$shim$             &used in all feature sets               &parselmouth \cite{parselmouth}     &\cite{ lopes2022performance,harar2020towards,10.1109/ACCESS.2019.2938265,10.1109/ACCESS.2018.2816338}\\
age                                                     &$age$              &used in all feature sets               &$-$                                & \cite{10.1109/ACCESS.2019.2938265}\\
\midrule
standard deviation of \gls{f0}                          &$\sigma_{\mathbf{f}_0}$&used / not used                &parselmouth \cite{parselmouth}, %
                                                                                                                 numpy \cite{2020NumPy-Array}       &\cite{lopes2022performance, harar2020towards}\\
occurrence of NaN values in \gls{f0}-related features   &$NaN$              &used / not used               &$-$                                & \textbf{Novel} \\                                                                                                                 
pitch difference      &$\Delta f_0$       &used / not used                    &parselmouth \cite{parselmouth}, %
                                                                                                                 numpy \cite{2020NumPy-Array}       &\textbf{Novel}\\
Shannon entropy                                         &$H$                &used / not used                    &spkit \cite{nikesh_bajaj_2021_4710694}&\cite{ junior2023multiple}\\
mean values of the first 20 \gls{lfcc}                  &$\overline{\mathbf{LFCC}}$&used / not used             &torchaudio \cite{torchaudio}       & \cite{reddy2021comparison}\\
mean values of the first three formants                 &$\overline{\mathbf{f}}$&used / not used                &parselmouth \cite{parselmouth}     &\cite{hemmerling2017voice}\\
skewness                         &$skew$&used / not used                                 &SciPy \cite{2020SciPy-NMeth}   &\cite{barreira2020kullback}\\
mean spectral centroid                                  &$\overline{S}$     &used / not used                    &librosa \cite{librosa},%
                                                                                                                 numpy \cite{2020NumPy-Array}       &\cite{Degila_Errattahi_Hannani_2018}\\
mean spectral contrast                                  &$\overline{\mathbf{SC}}$&used / not used               &librosa \cite{librosa}, %
                                                                                                                 numpy \cite{2020NumPy-Array}       &\cite{Degila_Errattahi_Hannani_2018}\\
mean spectral flatness                                  &$\overline{SF}$    &used / not used                    &librosa \cite{librosa}, %
                                                                                                                 numpy \cite{2020NumPy-Array}       &\cite{Parsa_Jamieson_2000}\\
mean spectral roll-off                                  &$\overline{RO}$    &used / not used                    &librosa \cite{librosa}, %
                                                                                                                 numpy \cite{2020NumPy-Array}       &\cite{Degila_Errattahi_Hannani_2018}\\
mean zero-crossing rate                                 &$\overline{ZCR}$   &used / not used                    &librosa \cite{librosa}, %
                                                                                                                 numpy \cite{2020NumPy-Array}       &\cite{junior2023multiple}\\
\midrule
mean values of the selected \gls{mfcc}                  &$\overline{\mathbf{MFCC}}$ & \multirow{3}{*}{not used / used 13 / used 20} &librosa \cite{librosa},%
                                                                                                                 numpy \cite{2020NumPy-Array}       &\cite{borsky2017modal, 10.1109/ICASSP48485.2024.10446075, 10.1016/j.jestch.2022.101148, 10.1109/ACCESS.2018.2816338, tirronen2023hierarchical,yagnavajjula2024automatic,fan2021class}\\
mean first derivative of the selected \gls{mfcc}        &$\overline{\Delta \mathbf{MFCC}}$& &librosa \cite{librosa}, numpy \cite{2020NumPy-Array}   &\cite{10.1109/ACCESS.2018.2816338,tirronen2023hierarchical, yagnavajjula2024automatic,fan2021class}\\
mean second derivative of the selected \gls{mfcc}       &$\overline{\Delta^2 \mathbf{MFCC}}$& &librosa \cite{librosa}, numpy \cite{2020NumPy-Array} &\cite{10.1109/ACCESS.2018.2816338,tirronen2023hierarchical, yagnavajjula2024automatic,fan2021class}\\
\midrule
variance of selected \gls{mfcc}                         &$\mathbf{\sigma^2_{MFCC}}$& \multirow{3}{*}{ used / not used } &librosa \cite{librosa},%
                                                                                                                 numpy \cite{2020NumPy-Array}       &\cite{harar2020towards, yagnavajjula2024automatic}\\
variance of first derivative of the selected \gls{mfcc} &$ \mathbf{\sigma^2_{\Delta MFCC}}$& &librosa \cite{librosa}, numpy \cite{2020NumPy-Array}%
                                                                                                                                            &\cite{yagnavajjula2024automatic}\\
variance of second derivative of the selected \gls{mfcc}&$ \mathbf{\sigma^2_{\Delta^2MFCC}}$& &librosa \cite{librosa}, numpy \cite{2020NumPy-Array}%
                                                                                                                                            &\cite{yagnavajjula2024automatic}\\
\bottomrule
\end{tabular}
\end{sidewaystable*}

\subsubsection{Pitch Difference} \label{subsec:pitch_difference}
To improve classification, we introduce a novel feature, the pitch difference.
We determine the pitch difference as the difference ($\Delta f_0$) between the maximum and minimum values relative to the minimum value of the extracted \gls{f0} (Equation~\ref{eq:diffpitch}). Our reasoning is that a healthy voice should maintain a stable frequency for the full duration of the recording compared to the pathological voice, which may fluctuate. Note, that this feature is dimensionless and allows for comparisons across different scales and dataset.

\begin{equation}
    \label{eq:diffpitch}
    \Delta f_0 =\cfrac{ \max (\mathbf{f}_0) - \min (\mathbf{f}_0)}{\min (\mathbf{f}_0)}
\end{equation}

\subsubsection{NaN Feature}

For eight recordings, specifically \textit{492-a\_n.wav, 719-a\_n.wav, 720-a\_n. wav}, \textit{915-a\_n.wav, 1338-a\_n.wav, 1407-a\_n.wav, 1716-a\_n.wav}, and \textit{2235-a\_n.wav}, the algorithm extracting the \gls{f0}, implemented in \textit{parselmouth} library, fails and therefore, the feature values for $\mu_{f_0}$, $\Delta f_0$, $\sigma_{f_0}$, $\text{jitta}$, and $\text{shim}$ are NaN. This is probably caused by the dominance of the disharmonic part for serious cases of voice pathology. We replace the NaN values with zero and add a new binary feature that takes the value of 1 for the occurrence of NaN values. As this inability to detect the fundamental frequency is an important information about the recording, we consider introducing this feature a legitimate approach. Note that these eight recordings represent 2 female and 6 male subjects in our dataset, and all of them suffer from some form of voice pathology according to the database records.

\subsubsection{Feature Sets}\label{subsec:feature_dataset}

\begin{figure}[!h]
    \centering
    \includegraphics[width=0.5\columnwidth]{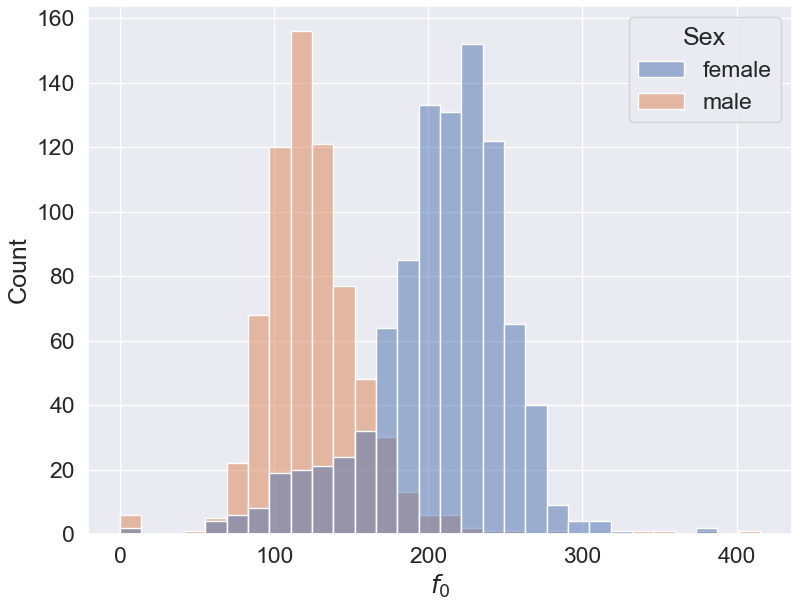}
    \caption{Distribution of the estimated mean \gls{f0} values among male and female patients}
    \label{fig:f0dist_all}
\end{figure}

In the extracted features, there are significant differences in some feature values between male and female patients. This is likely caused by the higher pitch of the female voice compared to the male voice, altering characteristics such as \gls{f0} (see Figure~\ref{fig:f0dist_all}). Since several features are dependent on \gls{f0}, even if indirectly, and the patient's biological sex is known during examination, we treat the data as two separate datasets and train two separate classification models, one for each sex. To verify this approach, we also train models for both sexes together.

We test different configurations of features to see their influence on classification performance. The possible configurations of optional features are listed in Table~\ref{tab:features_desc}. In total, we generate 20480 feature subsets, each for males, females, and both sexes. Note that exhaustive feature selection would lead to evaluation of more than $10^{49}$ feature sets, therefore, we include \gls{f0}, \gls{hnr}, jitter, shimmer and age in all generated feature sets. For \gls{mfcc} and their derivatives, we also limit the potential combinations to using no coefficients and derivatives, or first 13 and 20 coefficients and their derivatives, respectively. The combinations are then further expanded with the variances of the coefficients and their derivations.

\subsection{Data Augmentation}\label{sec:aug}
In order to address the issue of class imbalance in the utilized dataset, which leads to models with high recall and low specificity, we employ the k-means \gls{smote} algorithm \cite{kmeanssmote}. This technique is specifically applied to the training set to enhance the model's ability to learn from minority class instances and thus improve its predictive performance.

K-means \gls{smote} is an advanced oversampling method that combines the clustering capabilities of k-means \cite{lloyd1982least} with the synthetic data generation process of \gls{smote} \cite{chawla2002smote}.

The method has a risk of failing if the locations of the clusters are initialized close to outliers; therefore, we try initializing the method repeatedly, up to a maximum of ten times. If the method fails after ten repetitions, the algorithm is interrupted and \gls{smote} \cite{chawla2002smote} is used instead.

We apply the \gls{smote}-based algorithm only to the training data, to prevent data leakage. This approach allows us to mitigate the adverse effects of class imbalance, resulting in a more robust and reliable predictive model. 

\subsection{Machine Learning Algorithms}
\label{sec:model}
We perform binary classification using \gls{nb}, \gls{knn}, \gls{dt}, \gls{rf}, AdaBoost, and \gls{svm}. Given the relatively high dimensionality of the feature sets with a limited number of samples, deep learning models could be challenging to apply effectively. Therefore, we believe that using traditional \gls{ml} algorithms is a suitable approach and aligns with the findings from the existing research.

All evaluated \gls{ml} algorithms were implemented using the \textit{scikit-learn v1.5.2} library \cite{scikit-learn}. The list of hyperparameters tuned using grid search along with their corresponding values for each algorithm is presented in \autoref{tab:ml_param}. The naming of hyperparameters in the tables in the following text corresponds to the naming of function parameters in \textit{scikit-learn}.

\begin{table*}[h!]
    \centering
    \caption{\Gls{ml} algorithm hyperparameters for grid search}
    \label{tab:ml_param}
    \begin{tabular}{l l p{0.5\columnwidth}}
        \hline
        Algorithm & Hyperparameter& Tested values \\
        \hline
        \multirow{4}{*}{SVM} & kernel                   & "rbf", "poly" \\
                             & degree ("poly" only)     & 2, 3, 4, 5, 6 \\
                             & gamma                    & 0.5, 0.1, 0.05, 0.01, 0.005, 0.001, "auto" \\
                             & \multirow{2}{*}{C}       & 0.1, 0.5, 1, 5, 10, 50, 100, 500, 1000, 3000, 5000, 7000, 10000, 12000 \\
        \hline
        \multirow{3}{*}{KNN} & n\_neighbors             & 1, 3, 5, 7, 9, 11, 13, 15, 17, 19, 21, 23 \\
                             & p                        & 1, 2 \\
                             & weights                  & "uniform", "distance" \\
        \hline
        \multirow{4}{*}{DT} & criterion                 & "gini", "entropy", "log\_loss" \\
                            & splitter                  & "best", "random" \\
                            & min\_samples\_split       & 2, 3, 4, 5, 6, 7, 8, 9, 10 \\
                            & max\_features             & "sqrt", "log2" \\
        \hline
        \multirow{3}{*}{RF} & criterion & "gini"  \\
                            & min\_samples\_split       & 2, 3, 4, 5, 6 \\
                            & n\_estimators             & 50, 75, 100, 125, 150, 175 \\
                            & max\_features & "sqrt" \\
        \hline
        \multirow{2}{*}{AdaBoost} & learning\_rate      & 0.1, 1, 10 \\
                            & n\_estimators             & 50, 100, 150, 200, 250, 300, 350, 400 \\
        \hline
        NB                  & var\_smoothing            & 1e-8, 1e-9  \\
        \hline
    \end{tabular}
\end{table*}

\subsection{Validation of Results}
\label{sec:validation}
It is crucial to establish robust performance metrics to accurately assess model capability in distinguishing between healthy and pathological patients. Equally important is to ensure that the reported validation results are robust, minimizing the influence of randomness to guarantee the reliability and consistency of our conclusions, as well as the possibility to independently reproduce our findings.

\subsubsection{Used Metrics}
The imbalance between healthy and pathological samples in the dataset (407 healthy females, 541 females with pathologies, 252 healthy males and 436 males with pathologies) introduces bias into commonly used metrics such as accuracy, F1 score, precision, and negative predictive value \cite{imbalance_metrics}. Especially, accuracy can dangerously show overoptimistic results and provide misleading information \cite{chicco2020advantages}.

To address this issue and to provide metrics that reflect class imbalance, we evaluate the model performance using sensitivity (Equation~\ref{eq:sen}), specificity (Equation~\ref{eq:spe}), \gls{uar} (Equation~\ref{eq:un_rec}), \gls{gm} (Equation~\ref{eq:gm}), and \gls{bm} (Equation~\ref{eq:bm}). As all these metrics provide information about the successful classification, we also evaluate \gls{mcc} (Equation~\ref{eq:mcc}), which, although not entirely unbiased, has a smaller bias compared to accuracy and also takes misclassification into account \cite{imbalance_metrics}.

True positive ($\text{TP}$) predictions are correctly predicted positive (pathological) samples and true negative ($\text{TN}$) predictions are correctly predicted negative (healthy) samples. False positive ($\text{FP}$) predictions mark positive predictions of negative samples and false negative ($\text{FN}$) predictions mark negative predictions of positive samples.

\begin{equation}
    \label{eq:sen}
    \text{Sensitivity}=\frac{\text{TP}}{\text{TP}+\text{FN}}
\end{equation}

\begin{equation}
    \label{eq:spe}
    \text{Specificity} = \frac{\text{TN}}{\text{TN} + \text{FP}}
\end{equation}

\begin{equation}
    \label{eq:gm}
    \text{GM} = \sqrt{\text{Sensitivity} \cdot \text{Specificity}}
\end{equation}

\begin{equation}
    \label{eq:un_rec}
    \text{UAR} = \frac{\text{Sensitivity} + \text{Specificity}}{2}
\end{equation}

\begin{equation}
    \label{eq:bm}
    \text{BM} = \text{Sensitivity} + \text{Specificity} - 1
\end{equation}

\begin{equation}
    \label{eq:mcc}
    \text{MCC} = \cfrac{\text{TP}\cdot \text{TN} - \text{FP}\cdot \text{FN}}{\sqrt{\begin{aligned}(\text{TP}+\text{FP})\cdot(\text{TP}+\text{FN})\cdot\\ \cdot(\text{TN}+\text{FP})\cdot (\text{TN}+\text{FN})\end{aligned}}}
\end{equation}

Note that \gls{bm}, \gls{uar}, and \gls{gm} give the same weight to sensitivity and specificity, regardless of the distribution of positive and negative samples in the dataset. \gls{bm} and \gls{gm} penalize the performance on the minority class more compared to \gls{uar}, resulting in 0 values for 0 sensitivity (or specificity).

\subsubsection{Validation Approach}
Due to the limited size of the dataset and its imbalance, we employ stratified 10-fold cross-validation during the grid search, which should lead to less biased results of model performances \cite{kohavi1995study}. Only the training folds are augmented using the k-means SMOTE-based algorithm (see Section~\ref{sec:aug}) in each iteration. Then, each feature in the training set is scaled to an interval $[0,1]$ with the min-max scaler. The parameters obtained for the scaling of the training folds are used to scale the validation fold to avoid potential data leakage.

\begin{algorithm*}
\caption{Results validation}
\small
\label{alg:validation}
\begin{algorithmic}[1]
\For{each  \gls{ml} algorithm}
\State load results of all models
\State sort results according to MCC score
\State take 1000 results with highest MCC score
\For{each model with corresponding feature set from previous step}
\State i = 0
\While{i $<$ 100}
\State split the feature set randomly to 10 stratified folds
\For{each fold}
\State use this fold as a validation set and oversample rest of folds with k-means SMOTE algorithm
\State find a scaling parameters to scale each feature in training set to interval $[0,1]$
\State scale features in validation set using the scaling  parameters from previous step 
\State fit the model
\State  compute performance metrics 
\EndFor
\State i = i + 1
\EndWhile
\State compute average performance metrics from all repetitions of stratified cross-validation
\EndFor

\State select the model with corresponding feature set that has the highest average MCC as the best performing model
\EndFor
\end{algorithmic}
\end{algorithm*}

This process yields ten values for each performance metric. We then calculate the mean value of each metric across the folds, following standard cross-validation practices.

\begin{figure}[!h]
    \centering
    \includegraphics[width=0.6\columnwidth]{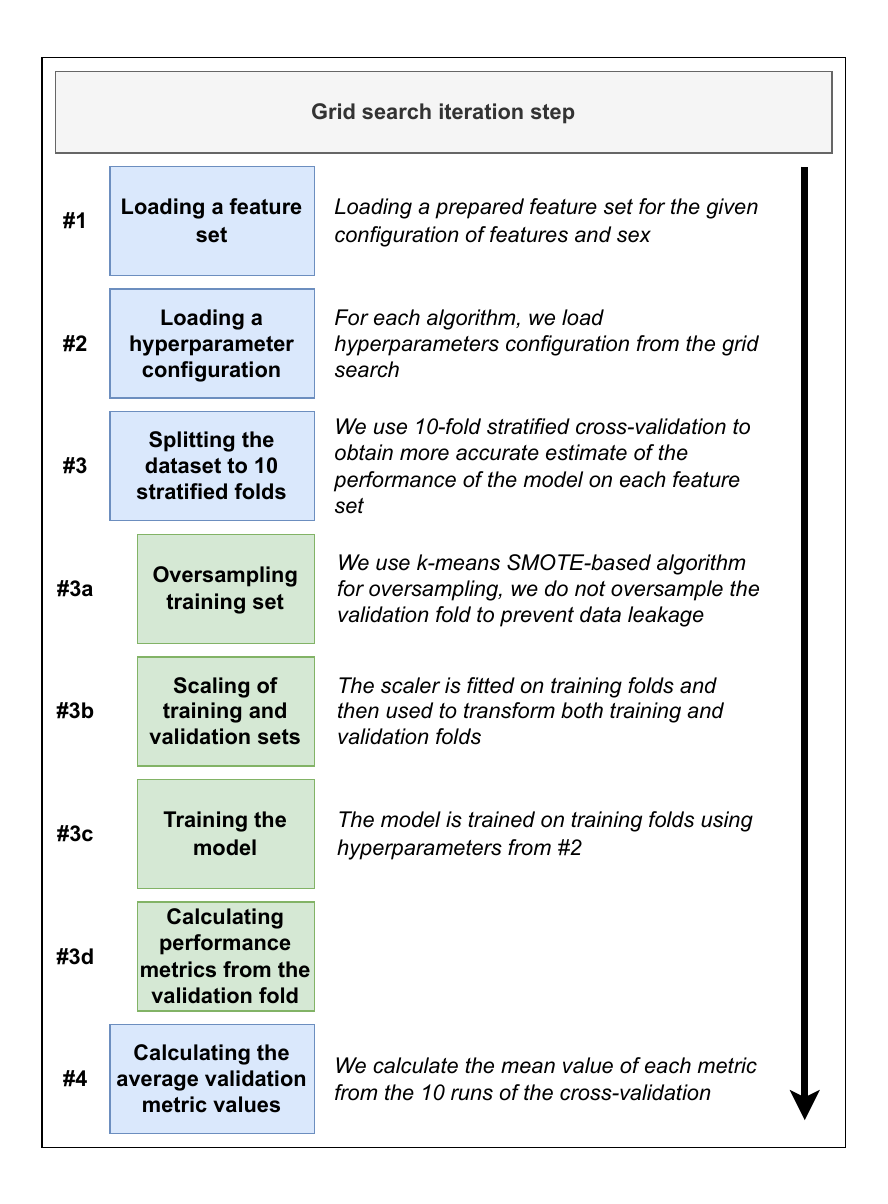}
    \caption{Single grid search iteration step}
    \label{fig:flow_pipeline}
\end{figure}

The workflow of the single grid search iteration step, from the augmentation to the calculation of the results from the cross-validation, is illustrated in Figure~\ref{fig:flow_pipeline}.

After completing the grid search, we identify the top 1000 performing models for each \gls{ml} algorithm based on \gls{mcc}. To account for the variance in \gls{mcc} introduced by the cross-validation splits, we perform 100 times repeated stratified 10-fold cross-validation for these models. This allows us to estimate the average metrics for each classification model and their corresponding standard deviations more reliably. The validation of the best results is shown in Algorithm~\ref{alg:validation}.

Moreover, we determine the top performing feature sets for each \gls{ml} algorithm. First, we aggregate the results from the grid search to get the average performance as well as the standard deviation for each feature set and \gls{ml} algorithm combination. Then, we select the best feature set for each \gls{ml} algorithm based on the highest value of \gls{mcc}.

\subsubsection{Ensuring Reproducibility}

To maintain the reproducibility of our findings, we implement multiple mechanisms so that all computations provide the same results and everyone can verify that our results were produced by the code we supply. The main problem arises from the unspecified license of the \gls{svd}. Therefore, we cannot share the feature sets, and we implement methods based on SHA256 checksums to validate input data and intermediate results. To make the use of methodology easier, we include a file named \verb|svd_information.csv| in the code repository, which contains metadata for each recording session, such as age, pathology, etc. The full list of files downloaded for this work is included in the repository along with SHA256 checksums.

Some algorithms used in the experiment were based on randomness, usually derived from the initial random values of trainable parameters. To mitigate these problems, we explicitly initialized \glspl{prng} with a seed (value 42 is used). We report exact versions of all software and computing hardware, and strict adherence to our setup is highly recommended.

Most computations are done in floating-point arithmetic, which introduces rounding errors. Moreover, to increase computational speed, software such as compilers does not always fully adhere to exact specifications and reference implementations. To ensure a way for other researchers to reproduce our results, we round some intermediate results where these errors happen, as well as increase floating point precision to minimize these errors.

\section{Results}

\label{sec:results}

All calculations are implemented in Python 3.12 \cite{10.5555/1593511}. The code is available at \url{https://github.com/aailab-uct/Automated-Robust-and-Reproducible-Voice-Pathology-Detection}. The \gls{ml} pipeline and k-means SMOTE-based algorithm are implemented using the \textit{imbalance-learn v0.12.4} library \cite{JMLR:v18:16-365}. The computations are done on multiple servers all with 2x AMD EPYC 9374F, 64 GB RAM with GNU/Linux OS Ubuntu 24.04 LTS. The libraries used for feature extraction, data augmentation, and model training are listed in the \textit{requirements.txt} file included in the aforementioned repository.

As we mentioned and explained in Section~\ref{sec:validation}, we do not present the accuracy score as the data used for training and evaluation of the dataset is moderately imbalanced. Instead, we provide an alternative in the form of \gls{mcc}, \gls{uar}, \gls{gm}, and \gls{bm}. However, if needed, accuracy can still be found in raw results reported in the repository.

\subsection{Top Performing Models}

The results for the best performing models (see step 20 of Algorithm~\ref{alg:validation}) for females, males and both sexes are presented in Table~\ref{tab:results_top_performance}.
The corresponding features and hyperparameter setting are in Table~\ref{tab:top_results_config_women} for females, in Table~\ref{tab:top_results_config_men} for males, and in Table~\ref{tab:top_results_config_both} for both sexes. Note, that a zero standard deviation for \gls{nb} corresponds to a subtle effect of Lidstone smoothing, which is manifested only in higher decimal places.

\begin{sidewaystable*}[ph!]
\small
\centering
\caption{Top model performance for each \gls{ml} algorithm}
\label{tab:results_top_performance}
\begin{tabular}{llcccccccccccc}\toprule\multirow{2}{*}{Sex} & \multirow{2}{*}{Algorithm} & \multicolumn{2}{c}{MCC} & \multicolumn{2}{c}{SEN} & \multicolumn{2}{c}{SPE} & \multicolumn{2}{c}{GM} & \multicolumn{2}{c}{UAR} & \multicolumn{2}{c}{BM} \\
& & $\mu$ & $\sigma$ & $\mu$ & $\sigma$ & $\mu$ & $\sigma$ & $\mu$ & $\sigma$ & $\mu$ & $\sigma$ & $\mu$ & $\sigma$ \\
\midrule
\multirow{6}{*}{F} & AdaBoost & \textbf{0.7157} & 0.0677 & 0.8942 & 0.0412 & 0.8152 & 0.0583 & 0.8530 & 0.0352 & 0.8547 & 0.0344 & 0.7094 & 0.0687\\
& DT & 0.6458 & 0.0755 & 0.8330 & 0.0537 & 0.8127 & 0.0627 & 0.8216 & 0.0384 & 0.8229 & 0.0379 & 0.6457 & 0.0758\\
& NB & 0.5402 & 0.0767 & 0.6905 & 0.0633 & \textbf{0.8511} & 0.0547 & 0.7652 & 0.0405 & 0.7708 & 0.0391 & 0.5416 & 0.0781\\
& KNN & 0.7063 & 0.0683 & 0.8762 & 0.0448 & 0.8273 & 0.0565 & 0.8506 & 0.0348 & 0.8518 & 0.0344 & 0.7036 & 0.0687\\
& RF & 0.7155 & 0.0681 & \textbf{0.8977} & 0.0411 & 0.8104 & 0.0578 & 0.8521 & 0.0354 & 0.8541 & 0.0345 & 0.7081 & 0.0689\\
& SVM & 0.7150 & 0.0670 & 0.8809 & 0.0424 & 0.8313 & 0.0558 & \textbf{0.8550} & 0.0342 & \textbf{0.8561} & 0.0338 & \textbf{0.7122} & 0.0675\\
\midrule
\multirow{6}{*}{M} & AdaBoost & 0.6375 & 0.0842 & 0.7941 & 0.0575 & 0.8608 & 0.0671 & 0.8255 & 0.0433 & 0.8274 & 0.0431 & 0.6549 & 0.0861\\
& DT & 0.5241 & 0.0957 & 0.7648 & 0.0640 & 0.7706 & 0.0845 & 0.7656 & 0.0499 & 0.7677 & 0.0491 & 0.5354 & 0.0982\\
& NB & 0.5438 & 0.0840 & 0.6673 & 0.0669 & \textbf{0.8928} & 0.0620 & 0.7703 & 0.0457 & 0.7801 & 0.0439 & 0.5601 & 0.0878\\
& KNN & 0.6216 & 0.0821 & 0.7637 & 0.0592 & 0.8778 & 0.0619 & 0.8175 & 0.0426 & 0.8207 & 0.0421 & 0.6414 & 0.0843\\
& RF & 0.6224 & 0.0912 & 0.8327 & 0.0533 & 0.7968 & 0.0809 & 0.8129 & 0.0479 & 0.8147 & 0.0468 & 0.6294 & 0.0937\\
& SVM & \textbf{0.6847} & 0.0853 & \textbf{0.8530} & 0.0500 & 0.8409 & 0.0736 & \textbf{0.8457} & 0.0441 & \textbf{0.8469} & 0.0436 & \textbf{0.6939} & 0.0873\\
\midrule
\multirow{6}{*}{B} & AdaBoost & 0.6509 & 0.0587 & 0.8392 & 0.0385 & 0.8154 & 0.0494 & 0.8265 & 0.0299 & 0.8273 & 0.0297 & 0.6545 & 0.0594\\
& DT & 0.5805 & 0.0627 & 0.7990 & 0.0431 & 0.7865 & 0.0517 & 0.7919 & 0.0319 & 0.7928 & 0.0316 & 0.5855 & 0.0633\\
& NB & 0.5075 & 0.0578 & 0.6256 & 0.0517 & \textbf{0.8839} & 0.0398 & 0.7427 & 0.0329 & 0.7547 & 0.0302 & 0.5095 & 0.0603\\
& KNN & 0.6583 & 0.0570 & 0.8120 & 0.0411 & 0.8552 & 0.0429 & 0.8328 & 0.0287 & 0.8336 & 0.0286 & 0.6673 & 0.0572\\
& RF & 0.6735 & 0.0565 & \textbf{0.8641} & 0.0352 & 0.8093 & 0.0489 & 0.8356 & 0.0291 & 0.8367 & 0.0286 & 0.6734 & 0.0572\\
& SVM & \textbf{0.6925} & 0.0556 & 0.8478 & 0.0366 & 0.8506 & 0.0432 & \textbf{0.8487} & 0.0279 & \textbf{0.8492} & 0.0278 & \textbf{0.6984} & 0.0557\\
\bottomrule
\end{tabular}
\end{sidewaystable*}

\begin{sidewaystable*}[ph!]
\centering
\caption{Top model -- feature set configuration for each \gls{ml} algorithm --- females}
\label{tab:top_results_config_women}
\begin{tabular}{lcccccc}
\toprule
Algorithm & \textbf{AdaBoost} & \gls{dt} & \gls{nb} & \gls{knn} & \gls{rf} & \gls{svm}\\
 & \shortstack[l]{'learning\_rate': \\\quad0.1\\ 'n\_estimators': \\\quad 300} & \shortstack[l]{'criterion': \\\quad 'gini'\\ 'max\_features':\\\quad'sqrt'\\ 'min\_samples\_split':\\\quad 10\\ 'splitter': \\\quad'random'} & \shortstack[l]{'var\_smoothing': \\\quad 1e-09} & \shortstack[l]{'n\_neighbors': \\\quad11\\ 'p':\\\quad2\\ 'weights': \\\quad'distance'} & \shortstack[l]{'criterion': \\\quad'gini'\\ 'max\_features': \\\quad 'sqrt'\\ 'min\_samples\_split':\\\quad 4\\ 'n\_estimators':\\\quad 175} & \shortstack[l]{'C': \\\quad 500\\ 'gamma': \\\quad0.5\\ 'kernel': \\\quad 'rbf'}\\
\midrule
$\overline{f}_{0}$ & Y & Y & Y & Y & Y & Y \\
$\overline{HNR}$ & Y & Y & Y & Y & Y & Y \\
$jitta$ & Y & Y & Y & Y & Y & Y \\
$shim$ & Y & Y & Y & Y & Y & Y \\
$NaN$ & Y & Y & N & N & Y & N \\
$age$ & Y & Y & Y & Y & Y & Y \\
$\sigma_{f_0}$ & Y & N & N & Y & Y & Y \\
$\Delta f_0$ & Y & Y & N & Y & Y & Y \\
$H$ & Y & N & Y & N & Y & N \\
$\overline{\mathbf{LFCC}}$ & Y & N & Y & N & N & N \\
$\overline{\mathbf{f}}$ & N & N & N & N & Y & Y \\
$skew$ & Y & N & N & N & Y & N \\
$\overline{S}$ & Y & N & N & N & N & N \\
$\overline{\textbf{SC}}$ & Y & N & N & N & N & N \\
$\overline{SF}$ & N & N & N & Y & Y & Y \\
$\overline{RO}$ & Y & N & N & N & N & N \\
$\overline{ZCR}$ & N & N & Y & N & Y & N \\
$\overline{\mathbf{MFCC}}$ & 20 & 0 & 0 & 0 & 0 & 0 \\
$\overline{\Delta \mathbf{MFCC}}$ & 20 & 0 & 0 & 0 & 0 & 0 \\
$\overline{\Delta^2 \mathbf{MFCC}}$ & 20 & 0 & 0 & 0 & 0 & 0 \\
$\mathbf{\sigma^2_{MFCC}}$ & 20 & N & N & N & N & N \\
$\mathbf{\sigma^2_{\Delta MFCC}}$ & 20 & N & N & N & N & N \\
$\mathbf{\sigma^2_{\Delta^2MFCC}}$ & 20 & N & N & N & N & N \\
\bottomrule
\end{tabular}
\end{sidewaystable*}

\begin{sidewaystable*}[ph!]
\centering
\caption{Top model -- feature set configuration for each \gls{ml} algorithm --- males}
\label{tab:top_results_config_men}
\begin{tabular}{lcccccc}
\toprule
{Algorithm} & AdaBoost & \gls{dt} & \gls{nb} & \gls{knn} & \gls{rf} & \textbf{\gls{svm}}\\
 & \shortstack[l]{'learning\_rate': \\\quad0.1\\ 'n\_estimators':\\\quad 400} & \shortstack[l]{'criterion':\\\quad 'entropy'\\ 'max\_features':\\\quad 'sqrt'\\ 'min\_samples\_split':\\\quad 9\\ 'splitter':\\\quad 'random'} & \shortstack[l]{'var\_smoothing':\\\quad 1e-09} & \shortstack[l]{'n\_neighbors':\\\quad 21\\ 'p':\\\quad 2\\ 'weights':\\\quad 'uniform'} & \shortstack[l]{'criterion':\\\quad 'gini'\\ 'max\_features':\\\quad 'sqrt'\\ 'min\_samples\_split':\\\quad 3\\ 'n\_estimators':\\\quad 175} & \shortstack[l]{'C':\\\quad 100\\ 'gamma': \\\quad0.05\\ 'kernel':\\\quad 'rbf'}\\
\midrule
$\overline{f}_{0}$ & Y & Y & Y & Y & Y & Y \\
$\overline{HNR}$ & Y & Y & Y & Y & Y & Y \\
$jitta$ & Y & Y & Y & Y & Y & Y \\
$shim$ & Y & Y & Y & Y & Y & Y \\
$NaN$ & N & N & N & N & N & N \\
$age$ & Y & Y & Y & Y & Y & Y \\
$\sigma_{f_0}$ & Y & N & N & N & Y & N \\
$\Delta f_0$ & N & N & N & N & Y & N \\
$H$ & N & N & N & N & Y & Y \\
$\overline{\mathbf{LFCC}}$ & Y & N & Y & N & N & Y \\
$\overline{\mathbf{f}}$ & N & N & Y & N & Y & N \\
$skew$ & Y & N & Y & N & N & N \\
$\overline{S}$ & Y & N & N & Y & Y & N \\
$\overline{\textbf{SC}}$ & Y & N & N & N & Y & Y \\
$\overline{SF}$ & N & N & N & N & N & Y \\
$\overline{RO}$ & Y & N & N & N & Y & N \\
$\overline{ZCR}$ & N & N & N & Y & N & N \\
$\overline{\mathbf{MFCC}}$ & 20 & 0 & 0 & 0 & 0 & 0 \\
$\overline{\Delta \mathbf{MFCC}}$ & 20 & 0 & 0 & 0 & 0 & 0 \\
$\overline{\Delta^2 \mathbf{MFCC}}$ & 20 & 0 & 0 & 0 & 0 & 0 \\
$\mathbf{\sigma^2_{MFCC}}$ & 20 & N & N & N & N & N \\
$\mathbf{\sigma^2_{\Delta MFCC}}$ & 20 & N & N & N & N & N \\
$\mathbf{\sigma^2_{\Delta^2MFCC}}$ & 20 & N & N & N & N & N \\
\bottomrule
\end{tabular}
\end{sidewaystable*}

\begin{sidewaystable*}[ph!]
\centering
\caption{Top model -- feature set configuration for each \gls{ml} algorithm --- both}
\label{tab:top_results_config_both}
\begin{tabular}{lcccccc}
\toprule
{Algorithm} & AdaBoost & \gls{dt} & \gls{nb} & \gls{knn} & \gls{rf} & \textbf{\gls{svm}}\\
 & \shortstack[l]{'learning\_rate': \\ \quad0.1\\ 'n\_estimators': \\ \quad400} & \shortstack[l]{'criterion': \\\quad 'gini'\\ 'max\_features': \\\quad'sqrt'\\ 'min\_samples\_split':\\\quad 10\\ 'splitter':\\\quad 'random'} & \shortstack[l]{'var\_smoothing': \\\quad 1e-08} & \shortstack[l]{'n\_neighbors': \\\quad 23\\ 'p':\\ \quad2\\ 'weights': \\\quad 'uniform'} & \shortstack[l]{'criterion': \\\quad 'gini'\\ 'max\_features': \\\quad'sqrt'\\ 'min\_samples\_split': \\ \quad6\\ 'n\_estimators': \\\quad175} & \shortstack[l]{'C': \\\quad 5\\ 'gamma': \\\quad 0.5\\ 'kernel': \\\quad 'rbf'}\\
\midrule
$\overline{f}_{0}$ & Y & Y & Y & Y & Y & Y \\
$\overline{HNR}$ & Y & Y & Y & Y & Y & Y \\
$jitta$ & Y & Y & Y & Y & Y & Y \\
$shim$ & Y & Y & Y & Y & Y & Y \\
$NaN$ & N & Y & N & N & N & N \\
$age$ & Y & Y & Y & Y & Y & Y \\
$\sigma_{f_0}$ & Y & N & N & Y & Y & Y \\
$\Delta f_0$ & N & N & N & Y & N & Y \\
$H$ & N & N & Y & N & N & Y \\
$\overline{\mathbf{LFCC}}$ & Y & N & Y & N & Y & Y \\
$\overline{\mathbf{f}}$ & N & N & N & N & Y & N \\
$skew$ & Y & N & Y & N & N & N \\
$\overline{S}$ & Y & N & N & Y & Y & N \\
$\overline{\textbf{SC}}$ & Y & N & N & N & N & N \\
$\overline{SF}$ & N & Y & N & Y & N & N \\
$\overline{RO}$ & N & N & N & N & N & N \\
$\overline{ZCR}$ & N & N & N & Y & N & Y \\
$\overline{\mathbf{MFCC}}$ & 20 & 0 & 0 & 0 & 0 & 0 \\
$\overline{\Delta \mathbf{MFCC}}$ & 20 & 0 & 0 & 0 & 0 & 0 \\
$\overline{\Delta^2 \mathbf{MFCC}}$ & 20 & 0 & 0 & 0 & 0 & 0 \\
$\mathbf{\sigma^2_{MFCC}}$ & N & N & N & N & N & N \\
$\mathbf{\sigma^2_{\Delta MFCC}}$ & N & N & N & N & N & N \\
$\mathbf{\sigma^2_{\Delta^2MFCC}}$ & N & N & N & N & N & N \\
\bottomrule
\end{tabular}
\end{sidewaystable*}

Based on the \gls{mcc} score, the best model for males is \gls{svm}. For the female dataset, the best performing model is AdaBoost, while the second best is \gls{rf}. It is worth mentioning that the \gls{svm} model performed in classification of female dataset just slightly worse than both AdaBoost and \gls{rf} model. For both sexes, the best performing model is \gls{svm}. The models for females perform better in general, which might be partially affected by the higher sample size and better ratio between the pathological and healthy samples. The top four female models outperformed the top male models and models including both sexes in \gls{mcc}.

Our proposed feature, the pitch difference, appears in the feature sets of the best performing models, being more prominent in the female feature sets. Specifically, it is utilized by the best performing female Ada\-Boost, \gls{rf}, \gls{svm}, and \gls{knn} models and by the best performing \gls{svm} model for both sexes. Regarding classification of males, the pitch difference was used only in \gls{rf} model, which was the third best model. The NaN feature was used by the best female models only, namely the AdaBoost and \gls{rf} models.

Feature sets with 13 \gls{mfcc} features were not present among the best performing models and 20 \gls{mfcc} features were used only by the female and male AdaBoost models. The variances of \gls{mfcc} were used by the male and female AdaBoost models. Notably, the top female AdaBoost models utilized almost complete feature set, where only formants, spectral flatness, and zero-crossing rate were omitted. Additionally, the entropy and \gls{lfcc} were used by the top female, male, and both-sex models. Generally, all features, except 13 \gls{mfcc}, were used at least by one top performing model.

\subsection{Top Performing Feature Sets}

\begin{table*}
\small
\centering
\caption{Top performing feature set for each \gls{ml} algorithm --- males}
\label{tab:best_datasets_men}

\begin{tabular}{{llcccccc}}
\toprule
\multicolumn{2}{l}{Algorithm} & \gls{svm} & \gls{knn} & \gls{nb} & \gls{dt} &\gls{rf} & AdaBoost \\
\midrule
\multicolumn{6}{l}{Performance metrics} \\
\multirow{2}{*}{MCC} & $\mu$ & 0.5765 & 0.5916 & 0.5440 & 0.4676 & \textbf{0.6384} & 0.6031 \\ 
 & $\sigma$ & 0.0488 & 0.0391 & 0.0000 & 0.0376 & 0.0109 & 0.0314 \\ 
\multirow{2}{*}{SEN} & $\mu$ & 0.7402 & 0.7598 & 0.6467 & 0.7626 & \textbf{0.8303} & 0.7730 \\ 
 & $\sigma$  & 0.0946 & 0.0085 & 0.0000 & 0.0195 & 0.0053 & 0.0554 \\ 
\multirow{2}{*}{SPE} & $\mu$ & 0.8469 & 0.8499 & \textbf{0.9125} & 0.7126 & 0.8179 & 0.8441 \\ 
 & $\sigma$ & 0.0724 & 0.0416 & 0.0000 & 0.0379 & 0.0096 & 0.0506 \\ 
\multirow{2}{*}{UAR} & $\mu$ & 0.7935 & 0.8049 & 0.7796 & 0.7376 & \textbf{0.8241} & 0.8085 \\ 
 & $\sigma$  & 0.0272 & 0.0210 & 0.0000 & 0.0199 & 0.0057 & 0.0160 \\ 
\multirow{2}{*}{GM} & $\mu$ & 0.7856 & 0.8021 & 0.7670 & 0.7351 & \textbf{0.8233} & 0.8045 \\ 
 & $\sigma$  & 0.0345 & 0.0207 & 0.0000 & 0.0206 & 0.0057 & 0.0176 \\ 
\multirow{2}{*}{BM} & $\mu$ & 0.5871 & 0.6097 & 0.5591 & 0.4753 & \textbf{0.6482} & 0.6170 \\ 
 & $\sigma$  & 0.0544 & 0.0420 & 0.0000 & 0.0397 & 0.0114 & 0.0320 \\ 
\midrule
\multicolumn{6}{l}{Features used in feature sets} \\
\multicolumn{2}{l}{$\overline{f}_{0}$} & Y & Y & Y & Y & Y & Y \\ 
\multicolumn{2}{l}{$\overline{HNR}$} & Y & Y & Y & Y & Y & Y \\ 
\multicolumn{2}{l}{$jitta$} & Y & Y & Y & Y & Y & Y \\ 
\multicolumn{2}{l}{$shim$} & Y & Y & Y & Y & Y & Y \\ 
\multicolumn{2}{l}{$NaN$} & N & Y & N & N & N & Y \\ 
\multicolumn{2}{l}{$age$} & Y & Y & Y & Y & Y & Y \\ 
\multicolumn{2}{l}{$\sigma_{f_0}$} & Y & N & N & N & N & N \\ 
\multicolumn{2}{l}{$\Delta f_0$} & Y & N & N & N & Y & Y \\ 
\multicolumn{2}{l}{$H$} & Y & N & Y & N & N & N \\ 
\multicolumn{2}{l}{$\overline{\mathbf{LFCC}}$} & N & N & N & N & N & Y \\ 
\multicolumn{2}{l}{$\overline{\mathbf{f}}$} & Y & N & N & N & Y & N \\ 
\multicolumn{2}{l}{$skew$} & N & N & Y & N & N & Y \\ 
\multicolumn{2}{l}{$\overline{S}$} & N & Y & N & N & Y & N \\ 
\multicolumn{2}{l}{$\overline{\textbf{SC}}$} & Y & N & N & N & Y & N \\ 
\multicolumn{2}{l}{$\overline{SF}$} & N & Y & N & N & N & Y \\ 
\multicolumn{2}{l}{$\overline{RO}$} & N & N & N & N & N & N \\ 
\multicolumn{2}{l}{$\overline{ZCR}$} & N & Y & N & N & Y & Y \\ 
\multicolumn{2}{l}{$\overline{\mathbf{MFCC}}$} & 20 & N & N & N & N & 20 \\ 
\multicolumn{2}{l}{$\overline{\Delta \mathbf{MFCC}}$} & 20 & N & N & N & N & 20 \\ 
\multicolumn{2}{l}{$\overline{\Delta^2 \mathbf{MFCC}}$} & 20 & N & N & N & N & 20 \\ 
\multicolumn{2}{l}{$\mathbf{\sigma^2_{MFCC}}$} & N & N & N & N & N & N \\ 
\multicolumn{2}{l}{$ \mathbf{\sigma^2_{\Delta MFCC}}$} & N & N & N & N & N & N \\ 
\multicolumn{2}{l}{$ \mathbf{\sigma^2_{\Delta^2MFCC}}$} & N & N & N & N & N & N \\ 
\bottomrule
\end{tabular}
\end{table*}

\begin{table*}
\small
\centering
\caption{{Top performing feature set for each \gls{ml} algorithm --- females}}
\label{tab:best_datasets_women}

\centering
\begin{tabular}{{llcccccc}}
\toprule
\multicolumn{2}{l}{{Algorithm}} & \gls{svm} & \gls{knn} & \gls{nb} & \gls{dt} &\gls{rf} & AdaBoost \\
\midrule
\multicolumn{6}{l}{Performance metrics} \\
\multirow{2}{*}{MCC} & $\mu$ & 0.5722 & 0.6842 & 0.5391 & 0.5772 & \textbf{0.7225} & 0.7196 \\ 
 & $\sigma$ & 0.0965 & 0.0391 & 0.0000 & 0.0288 & 0.0074 & 0.0030 \\ 
\multirow{2}{*}{SEN} & $\mu$ & 0.7366 & 0.8582 & 0.6710 & 0.7951 & 0.9055 & \textbf{0.9070} \\ 
 & $\sigma$  & 0.1182 & 0.0227 & 0.0000 & 0.0218 & 0.0050 & 0.0031 \\ 
\multirow{2}{*}{SPE} & $\mu$ & 0.8359 & 0.8251 & \textbf{0.8670} & 0.7831 & 0.8076 & 0.8030 \\ 
 & $\sigma$ & 0.0666 & 0.0173 & 0.0000 & 0.0210 & 0.0028 & 0.0024 \\ 
\multirow{2}{*}{UAR} & $\mu$ & 0.7862 & 0.8416 & 0.7690 & 0.7891 & \textbf{0.8565} & 0.8550 \\ 
 & $\sigma$  & 0.0495 & 0.0191 & 0.0000 & 0.0143 & 0.0035 & 0.0013 \\ 
\multirow{2}{*}{GM} & $\mu$ & 0.7575 & 0.8407 & 0.7611 & 0.7877 & \textbf{0.8544} & 0.8529 \\ 
 & $\sigma$  & 0.0880 & 0.0191 & 0.0000 & 0.0145 & 0.0035 & 0.0013 \\ 
\multirow{2}{*}{BM} & $\mu$ & 0.5724 & 0.6833 & 0.5380 & 0.5782 & \textbf{0.7131} & 0.7100 \\ 
 & $\sigma$  & 0.0990 & 0.0383 & 0.0000 & 0.0286 & 0.0070 & 0.0027 \\ 
\midrule
\multicolumn{6}{l}{Features used in feature sets} \\
\multicolumn{2}{l}{$\overline{f}_{0}$} & Y & Y & Y & Y & Y & Y \\ 
\multicolumn{2}{l}{$\overline{HNR}$} & Y & Y & Y & Y & Y & Y \\ 
\multicolumn{2}{l}{$jitta$} & Y & Y & Y & Y & Y & Y \\ 
\multicolumn{2}{l}{$shim$} & Y & Y & Y & Y & Y & Y \\ 
\multicolumn{2}{l}{$NaN$} & N & Y & N & Y & Y & Y \\ 
\multicolumn{2}{l}{$age$} & Y & Y & Y & Y & Y & Y \\ 
\multicolumn{2}{l}{$\sigma_{f_0}$} & Y & Y & N & Y & N & Y \\ 
\multicolumn{2}{l}{$\Delta f_0$} & Y & Y & N & N & Y & Y \\ 
\multicolumn{2}{l}{$H$} & Y & N & Y & N & Y & N \\ 
\multicolumn{2}{l}{$\overline{\mathbf{LFCC}}$} & N & N & N & N & N & N \\ 
\multicolumn{2}{l}{$\overline{\mathbf{f}}$} & N & N & Y & N & N & N \\ 
\multicolumn{2}{l}{$skew$} & Y & N & Y & N & Y & N \\ 
\multicolumn{2}{l}{$\overline{S}$} & N & N & N & N & Y & N \\ 
\multicolumn{2}{l}{$\overline{\textbf{SC}}$} & Y & N & N & N & N & Y \\ 
\multicolumn{2}{l}{$\overline{SF}$} & N & Y & N & N & N & N \\ 
\multicolumn{2}{l}{$\overline{RO}$} & Y & N & N & N & N & N \\ 
\multicolumn{2}{l}{$\overline{ZCR}$} & N & N & Y & N & N & Y \\ 
\multicolumn{2}{l}{$\overline{\mathbf{MFCC}}$} & 13 & N & N & N & N & N \\ 
\multicolumn{2}{l}{$\overline{\Delta \mathbf{MFCC}}$} & 13 & N & N & N & N & N \\ 
\multicolumn{2}{l}{$\overline{\Delta^2 \mathbf{MFCC}}$} & 13 & N & N & N & N & N \\ 
\multicolumn{2}{l}{$\mathbf{\sigma^2_{MFCC}}$} & N & N & N & N & N & N \\ 
\multicolumn{2}{l}{$ \mathbf{\sigma^2_{\Delta MFCC}}$} & N & N & N & N & N & N \\ 
\multicolumn{2}{l}{$ \mathbf{\sigma^2_{\Delta^2MFCC}}$} & N & N & N & N & N & N \\ 
\bottomrule
\end{tabular}
\end{table*}

\begin{table*}
\small
\centering
\caption{{Top performing feature set for each \gls{ml} algorithm --- both}}
\label{tab:best_datasets_both}
\begin{tabular}{{llcccccc}}
\toprule
\multicolumn{2}{l}{{Algorithm}} & \gls{svm} & \gls{knn} & \gls{nb} & \gls{dt} &\gls{rf} & AdaBoost \\
\midrule
\multicolumn{6}{l}{Performance metrics} \\
\multirow{2}{*}{MCC} & $\mu$ & 0.5936 & 0.6370 & 0.5140 & 0.5198 & \textbf{0.6745} & 0.6415 \\ 
 & $\sigma$ & 0.0692 & 0.0339 & 0.0000 & 0.0242 & 0.0093 & 0.0175 \\ 
\multirow{2}{*}{SEN} & $\mu$ & 0.7514 & 0.8160 & 0.6203 & 0.7793 & \textbf{0.8582} & 0.8343 \\ 
 & $\sigma$  & 0.1121 & 0.0112 & 0.0000 & 0.0151 & 0.0053 & 0.0099 \\ 
\multirow{2}{*}{SPE} & $\mu$ & 0.8428 & 0.8270 & \textbf{0.8938} & 0.7438 & 0.8172 & 0.8099 \\ 
 & $\sigma$ & 0.0700 & 0.0274 & 0.0000 & 0.0187 & 0.0068 & 0.0078 \\ 
\multirow{2}{*}{UAR} & $\mu$ & 0.7971 & 0.8215 & 0.7571 & 0.7616 & \textbf{0.8377} & 0.8221 \\ 
 & $\sigma$  & 0.0383 & 0.0175 & 0.0000 & 0.0122 & 0.0047 & 0.0083 \\ 
\multirow{2}{*}{GM} & $\mu$ & 0.7897 & 0.8203 & 0.7434 & 0.7602 & \textbf{0.8365} & 0.8210 \\ 
 & $\sigma$  & 0.0471 & 0.0174 & 0.0000 & 0.0124 & 0.0047 & 0.0080 \\ 
\multirow{2}{*}{BM} & $\mu$ & 0.5942 & 0.6430 & 0.5141 & 0.5231 & \textbf{0.6754} & 0.6442 \\ 
 & $\sigma$  & 0.0765 & 0.0349 & 0.0000 & 0.0244 & 0.0093 & 0.0165 \\ 
\midrule
\multicolumn{6}{l}{Features used in feature sets} \\
\multicolumn{2}{l}{$\overline{f}_{0}$} & Y & Y & Y & Y & Y & Y \\ 
\multicolumn{2}{l}{$\overline{HNR}$} & Y & Y & Y & Y & Y & Y \\ 
\multicolumn{2}{l}{$jitta$} & Y & Y & Y & Y & Y & Y \\ 
\multicolumn{2}{l}{$shim$} & Y & Y & Y & Y & Y & Y \\ 
\multicolumn{2}{l}{$NaN$} & N & Y & N & Y & N & Y \\ 
\multicolumn{2}{l}{$age$} & Y & Y & Y & Y & Y & Y \\ 
\multicolumn{2}{l}{$\sigma_{f_0}$} & N & Y & N & Y & Y & Y \\ 
\multicolumn{2}{l}{$\Delta f_0$} & Y & N & N & N & N & Y \\ 
\multicolumn{2}{l}{$H$} & Y & N & Y & N & Y & N \\ 
\multicolumn{2}{l}{$\overline{\mathbf{LFCC}}$} & Y & N & Y & N & Y & N \\ 
\multicolumn{2}{l}{$\overline{\mathbf{f}}$} & N & N & N & N & Y & N \\ 
\multicolumn{2}{l}{$skew$} & Y & N & N & N & Y & Y \\ 
\multicolumn{2}{l}{$\overline{S}$} & N & N & N & N & Y & Y \\ 
\multicolumn{2}{l}{$\overline{\textbf{SC}}$} & Y & N & N & N & N & Y \\ 
\multicolumn{2}{l}{$\overline{SF}$} & N & N & N & N & Y & Y \\ 
\multicolumn{2}{l}{$\overline{RO}$} & Y & N & N & N & Y & N \\ 
\multicolumn{2}{l}{$\overline{ZCR}$} & N & Y & Y & Y & N & Y \\ 
\multicolumn{2}{l}{$\overline{\mathbf{MFCC}}$} & 20 & N & N & N & N & N \\ 
\multicolumn{2}{l}{$\overline{\Delta \mathbf{MFCC}}$} & 20 & N & N & N & N & N \\ 
\multicolumn{2}{l}{$\overline{\Delta^2 \mathbf{MFCC}}$} & 20 & N & N & N & N & N \\ 
\multicolumn{2}{l}{$\mathbf{\sigma^2_{MFCC}}$} & N & N & N & N & N & N \\ 
\multicolumn{2}{l}{$ \mathbf{\sigma^2_{\Delta MFCC}}$} & N & N & N & N & N & N \\ 
\multicolumn{2}{l}{$ \mathbf{\sigma^2_{\Delta^2MFCC}}$} & N & N & N & N & N & N \\ 
\bottomrule
\end{tabular}
\end{table*}

The best performing feature sets for each \gls{ml} algorithm are selected by averaging their \gls{mcc} across all hyperparameter configurations (see Section~\ref{sec:model}). The best performing feature sets in combination with each \gls{ml} algorithm type are in Table~\ref{tab:best_datasets_women} for females, in Tables~\ref{tab:best_datasets_men} for males, and in Table~\ref{tab:best_datasets_both} for both sexes. 

Note that for each \gls{ml} algorithm, a different number of models was trained due to the different number of possible hyperparameter combinations. Moreover, the training was conducted on a single 10-fold cross-validation split which has a different influence on the final performance of each \gls{ml} algorithm. Therefore, the performance between models is not comparable.

The results show that the variances of \gls{mfcc} were absent completely in the best performing feature sets. \gls{lfcc} were not used in female feature sets and spectral roll-off was not used in male feature sets. The remaining features, including pitch difference and NaN feature, were present at least once.

\section{Discussion}

\begin{table*}[t]
\small
\begin{threeparttable}
\centering
\caption{Estimated mean bias of each metric for {the best model} for each sex}
\label{tab:metric_bias}
\begin{tabular}{| l|| r | r || r | r || r |r |}
\hline
Metric &  Female & Bias  & Male & Bias & Both & Bias \\
\hline
Sensitivity & 0.8942 & 0.0000 & 0.8530 & 0.0000 & 0.8478 & 0.0000\\
Specificity & 0.8152 & 0.0000 & 0.8409 & 0.0000 & 0.8506 & 0.0000\\
Accuracy & 0.8603 & 0.0056 & 0.8486 & 0.0016 & 0.8489 & -0.0003\\
F1-score & 0.8796 & 0.0194 & 0.8771 & 0.0293 & 0.8702 & 0.0212\\
Precision & 0.8654 & 0.0367 & 0.9027 & 0.0599 & 0.8938 & 0.0436\\
GM & 0.8538 & 0.0000 & 0.8469 & 0.0000 & 0.8492 & 0.0000\\
UAR & 0.8547 & 0.0000 & 0.8469 & 0.0000 & 0.8492 & 0.0000\\
MCC (normalized) \tnote{3} & 0.8388 & -0.0170 & 0.8378 & -0.0091 & 0.8463 & -0.0029\\
\hline
\end{tabular}
\begin{tablenotes}
\item[3]{Normalized MCC is transformation of MCC to the interval [0,1]}
\end{tablenotes}
\end{threeparttable}
\end{table*}

To support our preference for \gls{mcc} instead of accuracy, we estimate the biases for each metric for the best model in \autoref{tab:metric_bias}, according to \cite{imbalance_metrics}.

Many studies reported in Related Works section exclude data based on individual pathologies. While this may improve model performance and even allow multimodal classification of individual pathologies, we strongly believe this approach actually limits the applicability. In clinical practice, these models could not be utilized unless the excluded pathologies were also excluded from the possible diagnoses for the examined patients, which may prove impractical for clinical use.

None of the investigated studies reported handling the potential data leakage stemming from training models on datasets containing multiple recordings from the same patients. As there is a possibility that models can learn patterns of individual patients, given enough input, we assume many results of the works in \gls{ml}-based voice pathology detection might be overestimated due to this error. Therefore, we applied a method to exclude duplicate data based on patient identity. Our proposed approach does not discriminate against either healthy or pathological data and does not lead to overoptimistic results.


Our result is the combination of the best sex-aware models with results weighted by the number of subject of each sex. To our best knowledge, this is the first paper on voice pathology detection combining \gls{svd} and \gls{ml} methods, that is fully reproducible and conforms to the REFORMS practices \cite{kapoor2024reforms}. We provide the filled in REFORMS checklist in  \ref{sec:reforms}.

All features, except the three indicated in this paragraph, are obtained from voice recordings and are widely used in the models for voice pathology detection (see subsection Section~\ref{sec:related}). We consider age legitimate as a feature, as other acoustic features depend on the age of the speaker --- for example, changes in speaking fundamental frequency with aging \cite{nishio2008changes}.
The pitch difference (see subsection Section~\ref{subsec:pitch_difference}) is our proposed feature indicating the change of fundamental frequency during the recording and is extracted in a similar way as other considered features. The NaN feature reflects the fact that it was not possible to estimate the fundamental frequency for the patient (as described in Subsection~\ref{subsec:feature_dataset}). As the fundamental frequency is considered one of the dominant features in voice pathology diagnosis, we regard NaN feature as a legitimate approach.

\subsection{Limitations}
We are aware of several limitations our work is subject to. First, our models were tested using \gls{svd} only. The used database does not fully reflect the general population, especially in the proportion of healthy and pathological voices. However, at this time, there is no other suitable database that would reflect the general population better. Despite the justification as the only viable source of data, we cannot extrapolate its performance outside of this dataset. As the data was recorded in a controlled environment, we can assume our models might not be able to perform as well with datasets that are recorded under different conditions.

Moreover, we limit our research only to individuals who are 18 years old and older. Another noteworthy limitation was the available computational capacity, which led to careful decision of the \gls{ml} algorithms, hyperparameter space, and selected features we drew on throughout our work.

We also acknowledge that many voice and speech disorders extend beyond phonatory function, affecting articulation, prosody, and temporal dynamics, which are best analyzed through continuous speech rather than isolated vowel production. Neuromotoric diseases often present with abnormalities in articulation rate, syllable duration, coarticulation patterns, and prosodic modulation. These features cannot be fully evaluated through sustained vowel phonation alone, as they require an analysis of connected speech, where natural variations in stress, intonation, and fluency become more apparent \cite{Rong202492, Gómez2021} in the context of the four-level model and its implications for understanding the pathophysiology underlying apraxia of speech and other motor speech disorders \cite{Van_Der_Merwe2021397}.

In addition, deficits in speech motor control, imprecise consonant production, and irregularities in rhythm, which are key characteristics of certain neurological disorders, are best observed in spontaneous or structured speech tasks rather than in isolated vowel production. Including continuous speech in voice pathology detection frameworks enables a more comprehensive assessment of speech intelligibility, temporal variability, and segmental articulation, ultimately enhancing diagnostic accuracy and the clinical relevance of machine learning models. Acknowledging these limitations, future research should consider incorporating both sustained phonation and continuous speech samples into pathology detection systems to improve sensitivity to a wider range of voice and speech disorders.

\section{Conclusion}
\label{sec:conclusions}

In this study, we conducted an extensive comparison of \gls{ml}-based algorithms which are frequently used in related studies. We determined the best hyperparameter settings and feature combination by the grid search method. Moreover, we tested the influence of sex-based data split on the performance of the algorithms.


Next, we introduced several methodological concepts in the field of voice pathology detection and classification:
\begin{itemize}
    \item we provided reliable and reproducible results showing the top performance for models based on the selected \gls{ml} algorithms and hyperparameter settings, reporting realistic performance values,
    \item we introduced two novel features, pitch difference and NaN feature, which both were represented in the feature subsets that reached the reported best performance,
    \item we avoided potential introduction of data leakage by appropriate \gls{svd} data handling,
    \item we determined the performance by employing low-biased metrics such as \gls{mcc}, \gls{bm} and \gls{uar},
    \item we presented the top performing feature sets for each \gls{ml} algorithm with respect to all tested hyperparameter configurations.
\end{itemize}


Finally, there are several limitations to our work which are mostly based on a lack of suitable databases. Due to this fact, we omitted the external validation of the reached results.

\section*{Supplementary information}

\verb|list_of_excluded_files.csv|: recordings excluded along with reasons for exclusion (as described in Section~\ref{sec:data}).

\section*{Acknowledgements}

Jakub Steinbach and Tomáš Jirsa were supported by their specific university grant IGA A1\_FCHI\_2024\_002.

\section*{Declarations}

The authors declare that they have no known competing financial interests or personal relationships that could have appeared to influence the work reported in this paper.

\section*{Data availability}
All data used are publicly accessible on the following webpage: \url{https://stimmdb.coli.uni-saarland.de}

\section*{Code availability}
The code used to produce all results, along with all supplemental material and information to reproduce our results, is stored in publicly available GitHub repository \url{https://github.com/aailab-uct/Automated-Robust-and-Reproducible-Voice-Pathology-Detection} with the following DOI: \href{https://doi.org/10.5281/zenodo.13771573}{10.5281/zenodo.13771573}.

\section*{Author contribution}
\textbf{Jan Vrba:} Conceptualization, Formal analysis, Investigation, Methodology, Project administration, Software, Resources, Supervision, Visualization, Writing – original draft, Writing – review \& editing
\textbf{Jakub Steinbach:} Conceptualization, Data curation, Formal analysis, Funding acquisition, Investigation, Methodology, Software, Resources, Writing – original draft, Writing – review \& editing
\textbf{Tomáš Jirsa:}  Formal analysis, Data curation, Funding acquisition, Investigation, Methodology, Software, Resources, Validation, Writing – original draft, Writing – review \& editing
\textbf{Laura Verde:} Conceptualization, Formal analysis, Methodology, Software, Writing – original draft, Writing – review \& editing
\textbf{Roberta De Fazio:} Formal analysis, Software, Writing – review \& editing
\textbf{Zuzana Urbániová:} Formal analysis, Data curation, Validation, Writing – review \& editing
\textbf{Martin Chovanec:} Validation, Writing – review \& editing
\textbf{Yuwen Zeng:} Validation, Writing – review \& editing
\textbf{Kei Ichiji:} Validation, Writing – review \& editing
\textbf{Lukáš Hájek:} Resources, Validation, Writing – review \& editing
\textbf{Zuzana Sedláková:} Validation, Writing – review \& editing
\textbf{Jan Mareš:} Funding acquisition, Resources, Supervision, Writing – review \& editing
\textbf{Noriyasu Homma:} Methodology, Project administration, Supervision, Writing – review \& editing

\printglossaries

\appendix

\section{Reforms checklist}
\label{sec:reforms}

\input{appendix-reforms}

\bibliographystyle{ama} 
\bibliography{bibliography}






\end{document}

%% file: glossary.tex
\newacronym{dt}{DT}{decision trees} 
\newacronym{ai}{AI}{artificial intelligence} 
\newacronym{ml}{ML}{machine learning}
\newacronym{svd}{SVD}{Saarbrücken Voice Database} 
\newacronym{svm}{SVM}{support vector machine}
\newacronym{smote}{SMOTE}{synthetic minority over-sampling technique}
\newacronym{mfcc}{MFCC}{mel-frequency cepstral coefficients}
\newacronym{dnn}{DNN}{deep neural networks}
\newacronym{nn}{NN}{neural network}
\newacronym{cnn}{CNN}{convolutional neural network}
\newacronym{gfcc}{GFCC}{gammatone frequency cepstral coefficients}
\newacronym{lpc}{LPC}{linear predictive coefficients}
\newacronym{hnr}{HNR}{harmonic-to-noise ratio}
\newacronym{nb}{NB}{naive Bayes}
\newacronym{knn}{KNN}{k-nearest neighbors}
\newacronym{lstm}{LSTM}{long short-term memory}
\newacronym{egg}{EGG}{electroglottogram}
\newacronym{stft}{STFT}{short-time Fourier transform}
\newacronym{dct}{DCT}{discrete cosine transform}
\newacronym{lfcc}{LFCC}{linear-frequency cepstral coefficients}
\newacronym{f0}{\textbf{f$_0$}}{fundamental frequency}
\newacronym{zcr}{ZCR}{zero-crossing rate}
\newacronym{mlp}{MLP}{multi-layered perceptron}
\newacronym{meei}{MEEI}{Massachusetts Eye and Ear Infirmary}
\newacronym{avpd}{AVPD}{Arabic Voice Pathology Database}
\newacronym{femh}{FEMH}{Far Eastern Memorial Hospital}
\newacronym{uar}{UAR}{unweighted average recall}
\newacronym{voiced}{VOICED}{VOice ICar fEDerico II database}
\newacronym{rf}{RF}{random forest}
\newacronym{gm}{GM}{geometric mean}
\newacronym{bm}{BM}{bookmakers informedness}
\newacronym{mcc}{MCC}{Matthews correlation coefficient}
\newacronym{prng}{PRNG}{pseudo-random number generator}

%% file: appendix-reforms.tex
\subsection*{Module 1: Study design	}

\textbf{1a. State the population or distribution about which the scientific claim is made	}

	The population for our scientific claim consists of German females aged 18 to 94 and German males aged 18 to 89. These individuals recorded their voices pronouncing the /a:/ vowel at a normal pitch at the Institute für Phonetik, Universität des Saarlandes (see Table~\ref{tab:age_stats} and Figure~\ref{fig:age}).

\textbf{1b. Describe the motivation for choosing this population or distribution (1a).}

    In our research, we examine the feasibility of \gls{ml} methods for voice pathology detection in adult patients. The \gls{svd} is the only available suitable dataset, having a relatively large number of samples, while containing a wide number of diseases, which are also presented in the general population (see Section~\ref{sec:related} for reasoning and Section~\ref{sec:data} for description of the data).

\textbf{1c. Describe the motivation for the use of ML methods in the study.	}

    We aim to build models exploiting the large number of feature for automatic voice pathology detection system that maximize the Matthews correlation coefficient.

    In our research we do not investigate the relationship between feature causality or correlation with the pathology itself (see Section~\ref{sec:introduction}).

\subsection*{Module 2: Computational reproducibility	}
\textbf{2a. Describe the dataset used for training and evaluating the model and provide a link or DOI to uniquely identify the dataset.	}

    We utilize the publicly available dataset "Saarbruecken Voice Database" available at: \url{https://stimmdb.coli.uni-saarland.de/help_en.php4}. There is no unique identifier and no version control of the database, and we do not have a license to share the dataset, therefore, we provide list of files along with sha256 checksum to ensure reproducibility (see Section~\ref{sec:data}).

\textbf{2b. Provide details about the code used to train and evaluate the model and produce the results reported in the paper along with link or DOI to uniquely identify the version of the code used.}

    The code used to produce all results, along with all supplemental material and information to reproduce our results, is stored in publicly available GitHub repository \url{https://github.com/aailab-uct/Automated-Robust-and-Reproducible-Voice-Pathology-Detection} with following DOI: \href{https://doi.org/10.5281/zenodo.13771573}{10.5281/zenodo.13771573}.

\textbf{2c. Describe the computing infrastructure used.}

	All of the code produces the same results regardless of the architecture and operating system of the computer. The results were obtained on multiple servers with   2x AMD EPYC 9374F, 64 GB RAM with GNU/Linux OS Ubuntu 24.04 LTS (see Section~\ref{sec:results}).

\textbf{2d. Provide a README file which contains instructions for generating the results using the provided dataset and code.	}

	See the GitHub repository.

 \textbf{2e. Provide a reproduction script to produce all results reported in the paper.	}

	See the GitHub repository.	
 
\subsection*{Module 3: Data quality	}

\textbf{3a. Describe source(s) of data, separately for the training and evaluation datasets (if applicable), along with the time when the dataset(s) are collected, the source and process of ground-truth annotations, and other data documentation.	}

	All data are from Saarbruecken Voice Database as described in 2a. As this database is relatively small, we utilize only stratified 10-fold cross-validation without test or evaluation dataset. We downloaded data from \gls{svd} on July 12, 2022. The recordings used in our study were recorded between November 20, 1997 and June 16, 2004. The ground truth annotations were obtained by evaluation of stroboscopical recording by database authors \cite{svd}. All information related to \gls{svd} can be found at \url{https://stimmdb.coli.uni-saarland.de}.

\textbf{	3b. State the distribution or set from which the dataset is sampled (i.e., the sampling frame).	}

	As we only adapt the database, there is no information on the methodology of selection for the recording.

\textbf{3c. Justify why the dataset is useful for the modeling task at hand.	}

	We believe SVD dataset is relevant as it contains various pathologies that are also present in general population. In our study, after removing inappropriate recordings, we worked with 64 various pathologies (see Section~\ref{sec:data}).

 \textbf{3d. State the outcome variable of the model, along with descriptive statistics (split by class for a categorical outcome variable) and its definition.	}

	The outcome variable in this study is the health status of patients, classified into two categories: 'Healthy' and 'Pathological'. This binary classification is based on information provided by the database, specifically the information about pathologies.

    The outcome variable is derived from the table containing patient information. Classification takes the value of 0 if there are no values in the pathology column in the svd\_information.csv file, and the value of 1 if one or more pathologies are listed in the column (see Section~\ref{sec:data}). The distribution of healthy / pathological recordings of female and male subjects is provided in the Table~\ref{tab:age_stats}.	

\textbf{3e. State the sample size and outcome frequencies.}
 
	In our study, we utilized 1636 recordings out of 2041 total. In total, 977 are pathological and 659 are healthy. More detailed distribution is explained in Table~\ref{tab:age_stats}, Figure~\ref{fig:age}, and Figure~\ref{fig:flow_prep} (see Section~\ref{sec:data}).

\textbf{3f. State the percentage of missing data, split by class for a categorical outcome variable.}
 
	There was no missing data. However, 1 file (1573-a-n.wav) contains a recording of the time between two sessions and one is corrupted (87-a-n.wav). Moreover, there were multiple recordings marked as corrupt in comments to recording sessions. These recordings were excluded from our research. Finally, there were multiple recordings marked with pathologies "Gesangsstimme" and "S\"angerstimme", which probably contain healthy subjects. However, as it is not clear, we exclude this data unless they are diagnosed with another pathology (see Section~\ref{sec:data}).

\textbf{3g. Justify why the distribution or set from which the dataset is drawn (3b.) is representative of the one about which the scientific claim is being made (1a.).	}

	The used database does not fully reflect the general population, in the sense of proportion of healthy / pathological voices. However, at this time, there is no other suitable database that would reflect the general population better (see Section~\ref{sec:conclusions}).	
 
\subsection*{Module 4: Data preprocessing	}

\textbf{4a. Describe whether any samples are excluded with a rationale for why they are excluded.}

	From the dataset, we remove 13 corrupted recordings, 40 underage recordings, 17 singers and 335 recordings of patients with multiple recording sessions (except first healthy and first pathological recordings, see Figure~\ref{fig:flow_prep}). See Section~\ref{sec:data} for rationale and GitHub repository for list of excluded files (misc/ list\_of\_excluded\_files.csv).

\textbf{4b. Describe how impossible or corrupt samples are dealt with.	}

	When the extraction of \gls{f0} was impossible, the feature values for $\overline{f}_{0}$, $\sigma_{\mathbf{f}_0}$, $\Delta f_0$, jitta, and shim were set to 0 and the binary NaN feature was set to 1 (see Section~\ref{sec:feature}).

\textbf{4c. Describe all transformations of the dataset from its raw form (3a.) to the form used in the model, for instance, treatment of missing data and normalization—preferably through a flow chart.}
 
	See Figure~\ref{fig:flow_prep}. More details are in sections \ref{sec:data}, \ref{sec:feature}, and \ref{sec:aug}.	
 
\subsection*{Module 5: Modeling}

\textbf{5a. Describe, in detail, all models trained.}

	We utilize multiple ML algorithms for classification (see Table~\ref{tab:ml_param}) and the k-means SMOTE algorithm for dataset augmentation (see Section~\ref{sec:aug}).

\textbf{5b. Justify the choice of model types implemented.}

	All ML algorithms are suitable for multi-dimensional data, that we are dealing with (see Section~\ref{sec:model}).

\textbf{5c. Describe the method for evaluating the model(s) reported in the paper, including details of train-test splits or cross-validation folds.}

	Information about stratified 10-fold cross-validation and repeated stratified 10-fold cross-validation for the best models is described in the section Section~\ref{sec:validation}.

\textbf{	5d. Describe the method for selecting the model(s) reported in the paper.	}

	We performed repeated 10-fold cross-validation to estimate the average value of \gls{mcc} and its corresponding standard deviation. We select the best model according to this \gls{mcc}. See Section~\ref{sec:validation}.

\textbf{5e. For the model(s) reported in the paper, specify details about the hyperparameter tuning.}

	Hyperparameter tuning was approached via the grid search method. The range of hyperparameters was decided after preliminary experiments. See Table~\ref{tab:ml_param} for possible hyperparameter values.

\textbf{   5f. Justify that model comparisons are against appropriate baselines.}

	Our results are comparable to results in Section~\ref{sec:related}. Regarding reproducibility, we believe we are the first paper combining \gls{svd} and \gls{ml} methods while adhering to the REFORMS checklist. Our research distinguishes from the referred works by not eliminating data based on pathologies, by addressing potential data leakage through duplicities, by not oversampling on full dataset, and avoiding data lekage by improper data scaling. See more explanation in Section~\ref{sec:results}.
 
\subsection*{Module 6: Data leakage}

\textbf{6a. Justify that pre-processing (Module 4) and modeling (Module 5) steps only use information from the training dataset (and not the test dataset).	}

    By applying the oversampling algorithm only to the training folds, we aimed to prevent data leakage and ensure that the model's performance evaluation on the test fold remains unbiased. This approach allowed us to mitigate the adverse effects of class imbalance, resulting in a more robust and reliable predictive model. See Sections~\ref{sec:aug} and \ref{sec:validation}.

    The whole process, from preprocessing to validation, is described by Figures~\ref{fig:flow_prep}, \ref{fig:flow_pipeline}, and \ref{alg:validation}.

\textbf{6b. Describe methods used to address dependencies or duplicates between the training and test datasets (e.g. different samples from the same patients are kept in the same dataset partition).}

    For patients with multiple recordings of the same type (either all healthy or all pathological), we retained only the oldest recorded sample. For patients with both healthy and pathological recordings, we selected the oldest sample of each type, resulting in a maximum of two recordings per patient — one healthy and one pathological. We believe this approach minimizes the likelihood of the model learning patient identities, as the patient's classification remains independent of their identity. See Section~\ref{sec:data}.

\textbf{6c. Justify that each feature or input used in the model is legitimate for the task at hand and does not lead to leakage.}
 
	All features, except two, are obtained from voice recordings and are widely used in the models for voice pathology detection (see Section~\ref{sec:related}). We consider the "AGE" feature legitimate, as other acoustic features depend on the AGE. I.e. there are changes in speaking fundamental frequency with aging \cite{nishio2008changes}.

    The feature, that we introduced as "NaN" reflects the fact, that it was not possible to estimate the fundamental frequency for the patient. As the fundamental frequency is considered as one of the dominant features in voice pathology diagnosis, we consider introducing this "NaN" feature a legitimate approach. Note, that in total, there are 2 females and 6 males in our dataset, that have NaN value of fundamental frequency, all of them suffer from the voice disorder (see Section~\ref{sec:feature}).	

\subsection*{Module 7: Metrics and uncertainty}

\textbf{7a. State all metrics used to assess and compare model performance (e.g., accuracy, AUROC etc.). Justify that the metric used to select the final model is suitable for the task.}

	The choice of metrics, with respect to the class imbalance in the data, is written in the Section~\ref{sec:validation}. The claim regarding the best model is based on the Matthews correlation coefficient metric, that is suitable for imbalanced datasets and reflect both successes and errors in the classification.

\textbf{7b. State uncertainty estimates (e.g., confidence intervals, standard deviations), and give details of how these are calculated.}

	For each of metrics specified in Section~\ref{sec:validation}, we provide also the respective standard deviations that were obtained during the cross-validation procedure which is specified in this section.

\textbf{7c. Justify the choice of statistical tests (if used) and a check for the assumptions of the statistical test.}

	We do not use statistical tests in this study.	
 
\subsection*{Module 8: Generalizability and limitations}

\textbf{8a. Describe evidence of external validity.}

	As we consider \gls{svd} database for the only feasible database for our research, it is hard to describe evidence of external validity. See Section~\ref{sec:results}.

\textbf{8b. Describe contexts in which the authors do not expect the study’s findings to hold.}

	First, our model was tested using \gls{svd} only. The used database does not fully reflect the general population, in the sense of proportion of healthy / pathological voices. However, at this time, there is no other suitable database that would reflect the general population better. Despite the justification as an only viable source of data, we cannot extrapolate its performance outside of this dataset. Moreover, we limit our research only to individuals that are 18 years old and older. As the data was recorded in a controlled environment, we can assume our models might not be able to perform as well with datasets that are recorded during different conditions. Another noteworthy limitation was the available computational capacity which led to careful decision of the \gls{ml} algorithms and hyperparameter space we drew from during our work. See Section~\ref{sec:results}.